\documentclass[sigconf]{acmart}

\usepackage{bm}
\usepackage{amsmath}
\usepackage{enumitem}
\usepackage{graphicx}
\usepackage{amsfonts} 
\usepackage{multirow}
\usepackage{makecell}

\AtBeginDocument{%
  }

\copyrightyear{2025}
\acmYear{2025}
\setcopyright{acmlicensed}\acmConference[CIKM '25]{Proceedings of the 34th ACM International Conference on Information and Knowledge Management}{November 10--14, 2025}{Seoul, Republic of Korea}
\acmBooktitle{Proceedings of the 34th ACM International Conference on Information and Knowledge Management (CIKM '25), November 10--14, 2025, Seoul, Republic of Korea}
\acmDOI{10.1145/3746252.3761393}
\acmISBN{979-8-4007-2040-6/2025/11}

\begin{document}


\title{SUMMA: A Multimodal Large Language Model for Advertisement Summarization}

\author{Weitao Jia}
\email{jiaweitao@bytedance.com}
\affiliation{%
  \institution{ByteDance SearchAds}
  \city{Beijing}
  \country{China}
}

\author{Shuo Yin}
\email{yinshuo.01@bytedance.com}
\affiliation{%
  \institution{ByteDance SearchAds}
  \city{Beijing}
  \country{China}}

\author{Zhoufutu Wen}
\email{liniuniu@bytedance.com}
\affiliation{%
  \institution{ByteDance}
  \city{Beijing}
  \country{China}}

\author{Han Wang}
\email{wanghan.99@bytedance.com}
\affiliation{%
  \institution{ByteDance}
  \city{Beijing}
  \country{China}}

\author{Zehui Dai}
\email{zehuidai@bytedance.com}
\affiliation{%
  \institution{ByteDance SearchAds}
  \city{Beijing}
  \country{China}}

\author{Kun Zhang}
\email{zhangkun.9613@bytedance.com}
\affiliation{%
  \institution{ByteDance SearchAds}
  \city{Beijing}
  \country{China}}

\author{Zhenyu Li}
\email{lizhenyu.734@bytedance.com}
\affiliation{%
  \institution{ByteDance SearchAds}
  \city{Beijing}
  \country{China}}

\author{Tao Zeng}
\email{zengtao.7@bytedance.com}
\affiliation{%
  \institution{ByteDance SearchAds}
  \city{Beijing}
  \country{China}}

\author{Xiaohui Lv}
\authornote{Corresponding author.}
\email{lvxiaohui@bytedance.com}
\affiliation{%
  \institution{ByteDance SearchAds}
  \city{Beijing}
  \country{China}}


\renewcommand{\shortauthors}{Weitao Jia et al.}

\begin{abstract}
Understanding multimodal video ads is crucial for improving query-ad matching and relevance ranking on short video platforms, enhancing advertising effectiveness and user experience.
However, the effective utilization of multimodal information with high commercial value still largely constrained by reliance on highly compressed video embeddings—has long been inadequate.
To address this, we propose \textbf{SUMMA} (the abbreviation of \underline{\textbf{SU}mma}rizing \textbf{M}ulti\textbf{M}odal \textbf{A}ds), a multimodal model that automatically processes video ads into summaries highlighting the content of highest commercial value, thus improving their comprehension and ranking in our search-advertising systems.
SUMMA is developed via a two-stage training strategy—multimodal supervised fine-tuning followed by reinforcement learning with a mixed reward mechanism—on domain-specific data containing video frames and ASR/OCR transcripts, generating commercially valuable and explainable summaries.
We integrate SUMMA-generated summaries into our production pipeline, directly enhancing the candidate retrieval and relevance ranking stages in real search-advertising systems.
Both offline and online experiments show substantial improvements over baselines, with online results indicating a statistically significant 1.5\% increase in advertising revenue.
Our work establishes a novel paradigm for condensing multimodal information into representative texts, effectively aligning visual ad content with user query intent in retrieval and recommendation scenarios.
\end{abstract}






\maketitle

\section{Introduction}

\begin{figure*}
\includegraphics[width=\textwidth]{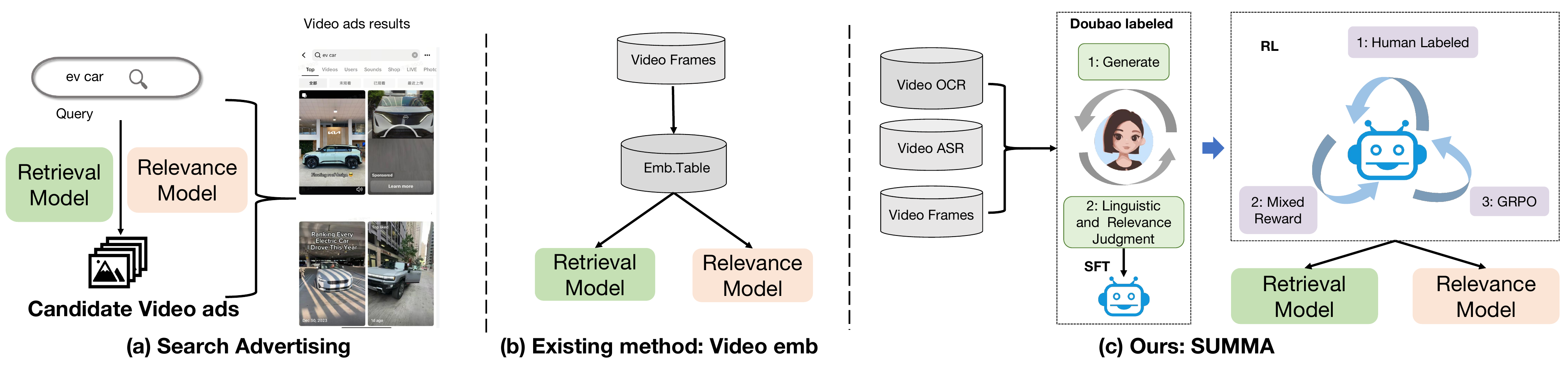}
\caption{(a) In search advertising, multimodal ad processing is key to optimal user-ad relevance. (b) Mainstream embedding-based methods rely on coarse multimodal fusion and lack interpretability. (c) Our SUMMA extracts video ad summaries, enabling text to replace original videos—easing fusion demands and boosting system performance. }
\end{figure*}

Short video platforms (e.g., Douyin, TikTok, Instagram Reels, and Kuaishou) significantly change how billions of people consume digital content every day. 
In this environment, search advertising becomes essential because video ads need to match user search intent accurately. Understanding multimodal content to identify valuable commercial information in ads is critical for effective advertising. 
However, current methods face challenges, especially in clearly defining what information has high commercial value—such as recognizing advertised products, brands, and key selling points—and efficiently extracting this information from videos.

Vision–text contrastive learning has surged in recent years, catalysing multimodal fusion research and greatly enhancing neural models’ capacity to predict over real-world visual–textual information.
\cite{CLIP,ALIGN,DeCLIP,ALBEF} show that large-scale and generic vision–language pre-training delivers strong cross-modal alignment, yet this paradigm largely ignores the domain-specific semantics, for business-oriented scenarios such as search advertising.
\cite{HCMRM,M2-RAAP,QUALITY,DQU-CIR,LDRE,vican} adapt these foundations to composed-image search, video–text retrieval and ad ranking, but the lack of interpretability and the essence of coarse global embeddings make them fail in fine-grained visual-textual correlations, thus degrading commercial reliability.
Recently, significant progress has been achieved in Large Language Models (LLMs)~\cite{InstructGPT,Llama2,Qwen2_5,Baichuan2,ChatGLM,InternLM2,DeepSeek-V2}, driving the substantial development of sophisticated Multimodal LLMs (MLLMs)~\cite{GPT-4,Llama-3,BLIP-2,LLaVA,MiniGPT-4,Qwen2.5-VL,InternVL2.5,DeepSeek-VL2,vora,elysium,videoqformer,dynamicvlm}, where modality alignment training among distinct modules leads to superior performance across diverse multimodal tasks, like visual question answering~\cite{MMBench,MMStar}, document understanding (OCR)~\cite{DocVQA,ChartQA,gloma}, and some vision-centric tasks~\cite{GLIP,CountBench}.
Despite their remarkable success, MLLMs remain underexplored in retrieval and recommendation systems. 

In this paper, we propose \textbf{SUMMA} (\underline{\textbf{SU}mma}rizing \textbf{M}ulti\textbf{M}odal \textbf{A}ds), a specialized MLLM for search advertising applications.
Initially, we perform supervised fine-tuning (SFT) from Qwen2-VL-2B-Instruct on an extensive collection of advertising short video summarization data (sourced from our short video platform). 
To ensure the quality of our SFT data, we implement a rule-based filtering approach leveraging LLM-as-a-judge.
This methodology employs dual criteria: assessing the intrinsic quality of video summaries and evaluating their potential to enhance performance in downstream relevance tasks.
By concatenating advertisement videos and their corresponding OCR/ASR texts as the input, the yielded SUMMA-SFT model exhibits a strong basic capacity for summarizing multimodal advertisements.
Subsequently, we further enhance SUMMA-SFT through reinforcement learning using high-quality meticulously annotated data. 
Our RL reward design draws inspiration from~\cite{MT-R1-Zero} and~\cite{BLEUBERI}, where the model receives encouragement or penalty signals based on various metrics with a reference (e.g., BLEU and external LLM judgment), depending on the generated outputs and human-annotated ground truth responses. 

The resulting model, called SUMMA-RL, demonstrates a significant improvement over baselines by generating concise yet informative summaries that efficiently process multimodal ad content into text of high commercial value, improving interpretability and staying friendly to latency-sensitive downstream advertising tasks.
Although reminiscent of image/video captioning, our \textbf{``visual ad summarization''} task uniquely produces representative textual content tailored to search-system ads, using a professional advertising lexicon to spotlight core semantic elements such as brands, selling points, and target audiences.

Extensive experimentation shows that combining video frames with OCR/ASR transcripts markedly outperforms any single-modality approach, and the margin widens as the textual stream becomes richer. 
Incorporating a mixed (joint lexical-and-semantic reward) reward mechanism further steers the model toward summaries that are simultaneously precise and meaning-preserving. 
These higher-quality summaries, in turn, yield clear downstream gains by raising retrieval hit rates and improving ranking AUC.
Overall, the key contributions of our work are as follows:
\begin{itemize}
\item We propose a data construction pipeline for the task of ad video summarization, where the generated video summary contains high commercial value by distilling key persuasive elements. 

\item We propose an efficient two-stage training paradigm, i.e., multimodal supervised fine-tuning followed by reinforcement learning, and experiments demonstrate that this scheme yields the best summarization quality and downstream performance.

\item 
We establish a search advertising framework taking SUMMA as the core multimodal comprehension component. 
The generated ad summaries enhance the downstream tasks--retrieval and relevance ranking--by improving query-ad alignment. Unlike methods relying solely on multimodal embeddings, our approach can better capture video semantics and key patterns, thus ultimately elevate the overall search advertising performance to a high level.

\item Both offline evaluation and online deployment in production environments consistently demonstrate the efficacy of the proposed SUMMA and its associated advertising architecture. 
The online experimental results demonstrate that our SUMMA totally yields a significant 1.5\% enhancement in advertising revenue.
Human evaluation further reveals a 5\% reduction in relevance bad case rate. 

\end{itemize}

\section{Related Work}
\subsection{Multimodal Approaches for Retrieval and Relevance}
Initial research on retrieval or recommendation systems mainly concentrate on unimodal textual analysis, thus limited in performance\cite{XMC,Que2Search,MASM,Pyramid-Ernie}. 
Fortunately, the evolution of cross-modal understanding has been significantly propelled by vision-language pretraining frameworks (e.g., contrastive learning based CLIP~\cite{CLIP}, as a seminal work in this field, ALIGN~\cite{ALIGN}, and DeCLIP~\cite{DeCLIP}), which bridge the semantic gap between visual and textual domains through large-scale corpus training. 
These models demonstrate remarkable transfer learning capabilities across diverse multimodal tasks. 
To align different modalities, Xu et al. propose ARMMT~\cite{ARMMT} to use cross-attention taking image and text embeddings as input, and output highly-fused multimodal representations. 
Similarly, based on cross attention, Gan et al.~\cite{HCMRM} design a pseudo-query-video matching task to further improve the modality aligning ability of their two-tower model.
For better processing the video information, Dong et al.~\cite{M2-RAAP} leverage temporal modeling modules (STAN~\cite{STAN}) in their zero-shot video-text retrieval pipeline and achieve notable effectiveness.
Consistent with human preference, Guo et al.~\cite{LR2PPO} utilize Proximal Policy Optimization (PPO~\cite{PPO}) to discern partial order relations among labels in multimodal label relevance ranking task.
Recently, Zhang et al.~\cite{NoteLLM-2} have proposed NoteLLM-2 which uses a multimodal understanding LLM to embed vision-text-mixed content into compressed representations. 
In contrast, our SUMMA opts to align with the pretrained pattern of MLLMs, generating high-quality texts being leveraged directly and naturally by the downstream retrieval and relevance ranking modules.

Considering the latency requirements for online services, the deployed downstream models need to be small-scale, and their inputs should also have limited length. However, directly using multimodal ad embeddings and user query text as inputs would impose an additional burden of multimodal fusion on such downstream task modules, which are inherently less capable due to their limited parameter amounts. 
On the contrary, the textual outputs from our SUMMA can be seamlessly integrated with downstream text-based similarity/relevance computation tasks, effectively alleviating the training burden and knowledge memorizing pressure on subsequent single-modal modules. 
Consequently, SUMMA allows downstream modules to focus exclusively on optimizing the alignment between user intent and advertising content.

\subsection{Multimodal Large Language Models}
Recent years have witnessed a surge in attention towards MLLMs, with their cross-modal understanding capabilities becoming a focal point in artificial intelligence research. 
Pioneering works like BLIP~\cite{BLIP}, Flamingo~\cite{Flamingo} and BLIP-2~\cite{BLIP-2} establish foundational architectures by integrating cross-modal attention mechanisms, demonstrating exceptional performance across diverse multimodal benchmarks. 
Following this trend, the community has focused on enhancing vision-language alignment through instruction training, exemplified by LLaVA~\cite{LLaVA}, MiniGPT-4~\cite{MiniGPT-4} and InstructBLIP~\cite{InstructBLIP}, which leverage large-scale image-text-mixed annotated/synthesized data to refine visual dialogue abilities. 
Among them, MiniGPT-4 and LLaVA have demonstrated that even simple modality fusion modules (like a linear projector) can achieve effective cross-modal alignment. 
Building upon this foundation, state-of-the-art frameworks such as LLaVA-NeXT~\cite{LLaVA-NeXT}, Qwen-VL series~\cite{Qwen-VL,Qwen2-VL,Qwen2.5-VL}, Intern-VL series~\cite{InternVL,InternVL1.5,InternVL2.5}, and DeepSeek-VL series~\cite{DeepSeek-VL,DeepSeek-VL2} support higher-resolution visual inputs for fine-grained visual cognition, via integrating enhanced visual encoders, image partitioning strategies, or sophisticated token compression algorithms. Moreover, these works also establish diverse and effective training paradigms to handle different intricate multimodal scenarios. 
Recently, a growing number of studies have focused on MLLMs for video understanding tasks~\cite{Video-LLaMA,VideoChat,Video-ChatGPT,InternVideo,InternVideo2,InternVideo2.5,ptseformer,dynamicvlm,videoqformer,elysium}, with particular emphasis on enhancing their spatiotemporal representation capabilities. 
Despite its great success, research on MLLMs remains underdeveloped in the domain of search advertising, resulting in constrained model capacities that fail to reach their full potential in this scenario.

Our proposed SUMMA, designed for video ad understanding, has undergone domain-specific fine-tuning and reinforcement learning on large-scale ad summarization datasets, where the OCR/ASR textual features supplementarily help enhance the overall performance. 
This specialized training pipeline enables effective deployment in search advertising downstream services such as retrieval and relevance ranking.

\subsection{Reinforcement Learning for MLLMs}
The recent OpenAI-o1~\cite{OpenAI_o1} and DeepSeek-R1~\cite{DeepSeek-R1} have catalyzed more and more interest within the open-source community in the enhancement of LLM reasoning capabilities through reinforcement learning (RL) training. 
Similarly, they have also stimulated research of applying RL to MLLMs.
Among them, Visual-RFT~\cite{Visual-RFT} outperforms conventional visual instruction tuning methods in computer vision tasks via an innovative reward mechanism depending on IoU or classification correction. 
Through Group Relative Policy Optimization (GRPO~\cite{DeepSeek-Math}), Visual-RFT demonstrates enhanced capabilities in object detection, visual grounding, and image recognition while requiring substantially less training data than SFT.
To boost RL policy performance, Skywork R1V~\cite{Skywork-R1V} firstly undergoes a cyclical error correction stage, where the model iteratively learns from previously mispredicted data instances. This curriculum learning paradigm, characterized by persistent refinement through challenging samples, demonstrates statistically significant improvements for efficacy of the subsequent reinforcement learning.
Moreover, Vision-R1~\cite{Vision-R1} adopts a phased RL approach of progressively increasing generation length constraints to enhance the improvements of their models. In contrast, Kimi-VL~\cite{Kimi-VL} implements a length penalty mechanism within its reward design, strategically preventing the model from producing excessively verbose responses.
Specially devised for video reasoning tasks, Video-R1~\cite{Video-R1} significantly enhances temporal reasoning capabilities by incorporating sequential temporal information into its reward mechanism design. This integrates the temporal dependency as a critical dimension during the decision-making process.

Similar to most of the aforementioned MLLM RL methods, our SUMMA also utilizes verifiable rewards, determined by the generated responses with reference to the ground truth answers. 
As for the differences, our task of interest is ad visual content summarization, and thus the suitable reward design is based on the lexical as well as semantic metrics against human-annotated references, inspired by~\cite{MT-R1-Zero,BLEUBERI}.

\begin{figure*}[ht]
  \centering
  \includegraphics[width=1.0\textwidth]{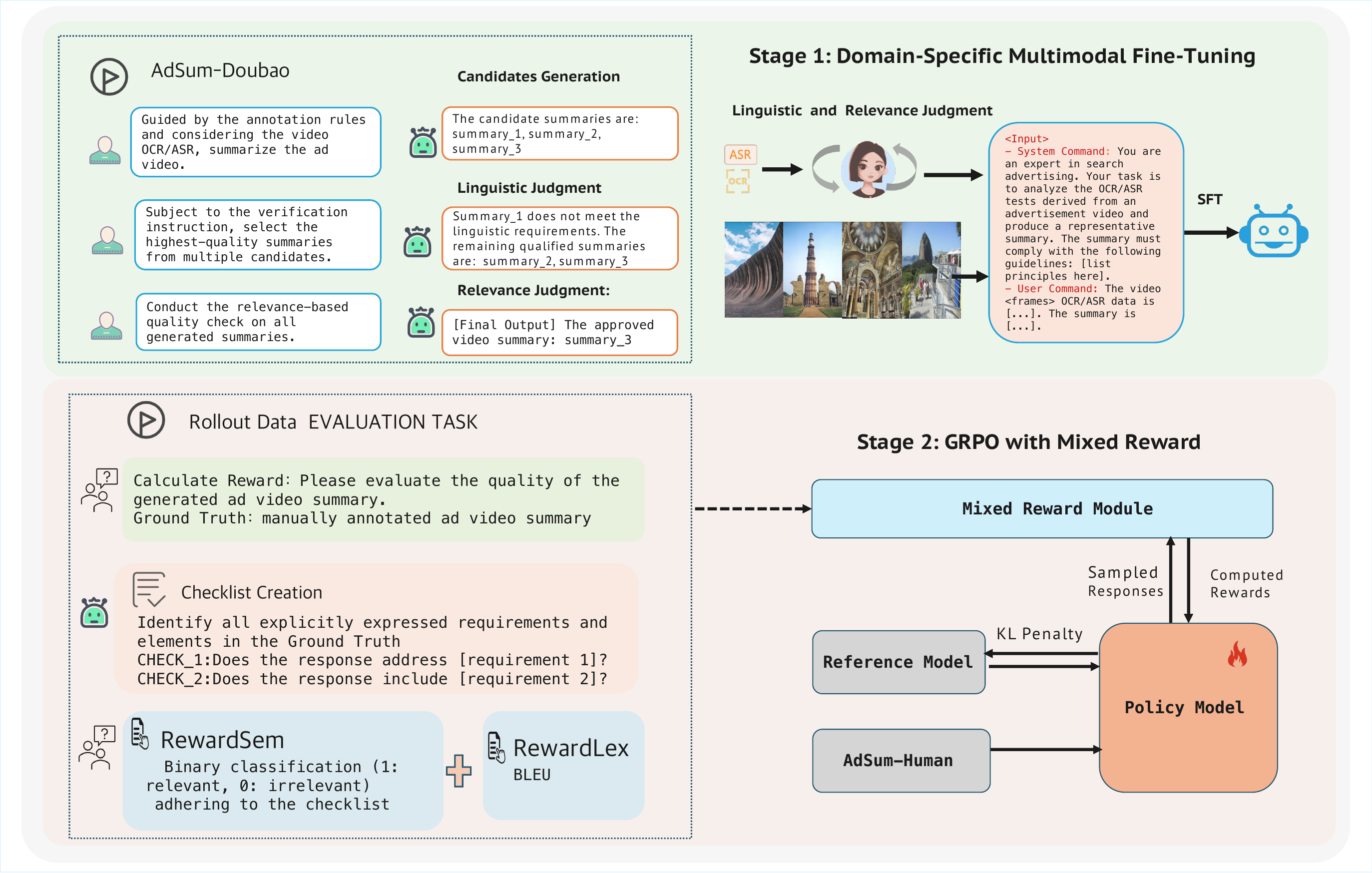}
  \caption{The overview pipeline of our SUMMA. In Stage 1, through cold-start SFT using ad summarization data, we develop a domain-specific MLLM, called SUMMA-SFT, which demonstrates fundamental capabilities in advertisement summarization. In Stage 2, building upon SUMMA-SFT, we implement reinforcement learning via GRPO with a mixed-reward mechanism, and the resultant model, SUMMA-RL, demonstrates further enhanced capacity.}
  \label{fig:method}
\end{figure*}

\section{Methodology}

In this section, we present \textbf{SUMMA}, an MLLM that 
generates representative, concise, and commercially oriented summaries by integrating advertising videos with their corresponding OCR/ASR transcripts.
As a roadmap for what follows, Figure~\ref{fig:method} illustrates the overall architecture of SUMMA. 
Building on the Qwen2-VL, the framework jointly encodes videos as well as the corresponding OCR/ASR transcripts, then decodes them into advertising-style highlights that emphasize product brands.
To optimize this ability, we adopt a two-stage pipeline.
1. Supervised Fine-Tuning (SFT) Stage:We perform data collection and processing by gathering advertising videos together with their OCR/ASR transcripts. Employing cold-start multimodal supervised fine-tuning with data generated and cleaned by Doubao-1.5-pro~\cite{Doubao-1.5-pro} (hereafter abbreviated as Doubao).
2. Reinforcement Learning (RL) Stage: Utilizing GRPO combined with a mixed-reward mechanism to further align the model with human preferences.
The remainder of this section details each component of the pipeline in turn.

\subsection{Data Curation}
\label{section:data_construction}

\begin{figure}[htbp]
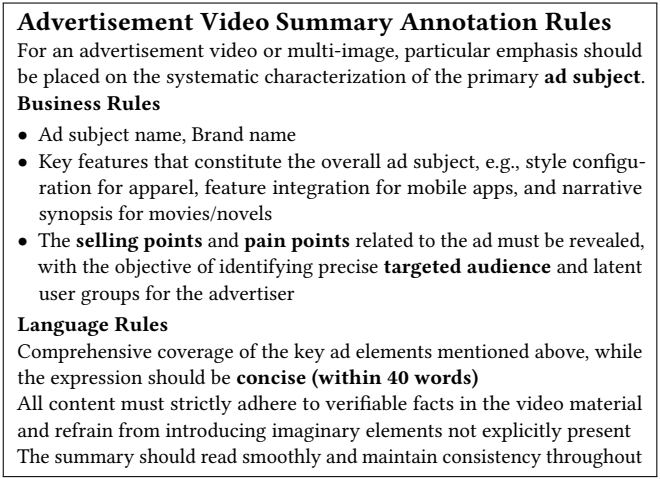

\centering
\fbox{
\begin{minipage}{0.98\linewidth}
\small
\textbf{\large{Advertisement Video Summary Annotation Rules}}\\
For an advertisement video or multi-image, particular emphasis should be placed on the systematic characterization of the primary \textbf{ad subject}. 
\textbf{Business Rules} 
\begin{itemize}[leftmargin=*,noitemsep]
\item Ad subject name, Brand name
\item Key features that constitute the overall ad subject, e.g., style configuration for apparel, feature integration for mobile apps, and narrative synopsis for movies/novels
\item The \textbf{selling points} and \textbf{pain points} related to the ad must be revealed, with the objective of identifying precise \textbf{targeted audience} and latent user groups for the advertiser
\end{itemize}
\textbf{Language Rules} \\
Comprehensive coverage of the key ad elements mentioned above, while the expression should be \textbf{concise (within 40 words)} \\
All content must strictly adhere to verifiable facts in the video material and refrain from introducing imaginary elements not explicitly present \\
The summary should read smoothly and maintain consistency throughout
\end{minipage}
}
\caption{The annotation rules of the multimodal ad content summarization task: both our experts and the AI tool should comply with these rules while summarizing an ad video.}
\label{fig:annotation_rules}
\end{figure}

\begin{figure}[t!]
\centering
\fbox{
\begin{minipage}{0.98\linewidth}
\small
\textbf{\large{Instruction of Summary Synthesis}}\\
You are an expert in search advertising. Your task is to analyze the OCR/ASR texts derived from an advertisement video and produce a distilled and representative summary that encapsulates the core message of the corresponding ad content. 
The summarization must adhere to the following operational principles: \textbf{\{annotation\_rules\}}

First, perform OCR/ASR analysis on the advertisement video to detect: (1) advertising subject presence, and (2) brand name identification. If both elements are present with adequate contextual information required in the above rules, proceed accordingly. Otherwise, return the standardized response: ``Insufficient information for summarization''.

\textit{Input Structure}
\begin{itemize}[leftmargin=*,noitemsep]
\item \texttt{ASR}: \textbf{\{asr\_text\}}
 , \texttt{OCR}: \textbf{\{ocr\_text\}}
\end{itemize}

\textit{Output Specification} 
\begin{itemize}[leftmargin=*,noitemsep]
\item \texttt{contain\_subject\_or\_not} : True/False
\item \texttt{subject\_name} : null if contain\_subject\_or\_not is False
\item \texttt{contain\_brand\_or\_not}: True/False
\item \texttt{brand\_name} : null if contain\_brand\_or\_not is False
\item \texttt{reason}: reason to judge if the summarization can be done
\item \texttt{summary}: ``Insufficient information for summarization'' if cannot be summarized
\end{itemize}

\end{minipage}
}
\caption{The prompt to request Doubao to synthesize ad video summaries.}
\label{fig:prompt_summary}
\end{figure}

\begin{figure}[t!]
\centering
\fbox{
\begin{minipage}{0.98\linewidth}
\small
\textbf{\large{Instruction for Summary Quality Verification}} \\
You are an expert in search advertising. Given the OCR/ASR texts derived from an advertisement video and the corresponding video summary generated by an AI assistant, your task is to conduct a systematic analysis of the summary's quality. This analysis should methodically evaluate the summary's compliance with each of the criteria outlined below:
\textbf{\{annotation\_rules\}}

After your analysis, you should make a final decision indicating the summary can pass your quality check or not.

\textit{Input Structure}
\begin{itemize}[leftmargin=*,noitemsep]
\item \texttt{ASR}: \textbf{\{asr\_text\}}
, \texttt{OCR}: \textbf{\{ocr\_text\}}
, \texttt{Summary}: \textbf{\{summary\_text\}}
\end{itemize}

\textit{Output Structure}
\begin{itemize}[leftmargin=*,noitemsep]
\item \texttt{reason} : reason to judge , \texttt{pass}: True/False 
\end{itemize}

\end{minipage}
}
\caption{The prompt to request Doubao to verify the quality of the synthesized ad video summaries.}
\label{fig:prompt_quality_verification}
\end{figure}

Leveraging our in-house search-ad video data, we curate three complementary resources—\textbf{AdSum-Doubao} (LLM-bootstrapped summaries for supervised fine-tuning, SFT), \textbf{AdSum-Human} (expert-annotated summaries for reinforcement learning, RL), and \textbf{AdSum-Test} (a held-out set reserved for final evaluation). 
The construction of each dataset is detailed in the following sections.

\paragraph{\textbf{AdSum-Doubao}}
Guided by the annotation scheme in Figure~\ref{fig:annotation_rules}, we prompt the Doubao LLM to produce \mbox{$\leq$40-word} synopses that highlight the ad subject, brand, and other commercially salient cues.  
Concretely, OCR/ASR transcripts extracted from each advertising video—together with basic metadata—are concatenated into the textual input to Doubao, following a modality-flattening strategy similar to LLaVA~\cite{LLaVA}.  
For every video we sample several candidate summaries by varying the temperature; a two-stage, Doubao-based automatic verifier then filters and ranks these outputs to yield the final dataset. \textbf{(1) Linguistic Judgment.} Using the validation criteria specified in Figure~\ref{fig:annotation_rules}, we select the highest-quality summary from multiple candidates. The prompt we utilize is shown in Figure~\ref{fig:prompt_quality_verification}. \textbf{(2) Relevance Judgment.}
Each sample of our relevance data includes a user query, a video and a human-annotated label (0/1). We feed the query and the relevance label, along with the video's synthesized summary, into Doubao for verification. This process determines whether each summary adequately reflects the relevance judgments (human feedback) of the corresponding query and ad. The LLM outputs a confidence score, only the summary with the highest score is preserved.

Consequently, only the summaries that pass both the linguistic-quality screening and the relevance filter are kept, leaving us with a clean and high-quality dataset for training.

\paragraph{\textbf{AdSum-Human}}
Domain experts first follow the protocol in Figure~\ref{fig:annotation_rules} to condense every ad video into a \mbox{$\leq$40-word} synopsis that foregrounds the ad subject, brand, and distinctive selling or pain points.  
The resulting 100K expert summaries then serve as the reward corpus in the RL stage, providing the high-confidence signals needed for stable policy optimisation.

\paragraph{\textbf{AdSum-Test}}
To evaluate the ad video summarization capabilities of our SUMMA, we collect 2.5k samples from our platform for manual annotation using the aforementioned methodology. 

\begin{table}[htbp]
\centering
\caption{Statistics of the three constructed datasets (AdSum-Doubao for SFT, AdSum-Human for RL, AdSum-Test for evaluation}
\resizebox{.35\textwidth}{!}{
\begin{tabular}{clc}
\toprule
\textbf{Dataset Name} & \textbf{Annotated} & \textbf{Size} \\ 
\hline
AdSum-Doubao & LLM & 400K \\
AdSum-Human & Human & 100K \\
AdSum-Test & Human & 2.5K \\ 
\bottomrule
\end{tabular}
}
\label{tab:dataset_stats}
\end{table}

\subsection{Domain-Specific Multimodal Fine-Tuning}

The initial fine-tuning phase imbues the model with domain-specific knowledge and processing capabilities. Formally, we optimize the following objective:
\begin{equation} 
\mathcal{L}_{\text{Domain-FT}}
= -\mathbb{E}_{(\mathbf{x_v}, \mathbf{x_{\text{aux}}}, \mathbf{y}) \sim \mathcal{D}_{\text{SFT}}} \left[ \log \pi(\mathbf{y} \mid \mathbf{x_v}, \mathbf{x_{\text{aux}}}; \bm\theta) \right],
\label{eq:domain-FT}
\end{equation}
where \(\mathbf{x_v},\mathbf{x_{\text{aux}}},\mathbf{y}\) are the video, auxiliary text (OCR/ASR) and summary, \(\mathcal{D}_{\text{SFT}}\) is the AdSum-Doubao dataset, and \(\pi(\cdot;\bm\theta)\) the multimodal transformer.

Upon completion of this domain-specific fine-tuning stage, we obtain the intermediate model \textbf{SUMMA-SFT}, which serves as the foundation for the subsequent RL stage.

\subsection{Mixed-Reward Reinforcement Fine-Tuning}
\label{subsection:admm_rl}

SUMMA-SFT effectively localizes key frames and extracts representative textual content tailored for ad recommendations.
However, this competence can be further improved through RL. Specifically, we employ GRPO to train our SUMMA-RL. 
For each query $\bm q = (\mathbf{x_v}, \mathbf{x_\text{aux}}) \sim \mathcal{D}_{RL}$, $N$ rollout responses are generated as a group, i.e., $\bm o^{(i)} \sim \pi_{\text{old}}(\bm o^{(i)} | \bm q)$ where $i = 1, 2, ..., N$, with corresponding rewards assigned are $\{r^{(1)}, r^{(2)}, \dots, r^{(N)}\}$. The optimization objective is defined as:

\begin{align}
\small
\mathcal{L}_{\text{GRPO}}
= \mathbb{E}_{\bm q,\bm o^{(i)}}\!\Bigl[
       & \min\!\bigl(
           \rho^{(i)}\hat A^{(i)},
           \operatorname{clip}\!\bigl(\rho^{(i)},1-\delta,1+\delta\bigr)\hat A^{(i)}
       \bigr)
       - \mathbb{D}_{\text{KL}}
   \Bigr] \normalsize \nonumber \\[6pt] 
\hat A^{(i)} &= \frac{r^{(i)} - \operatorname{mean}\!\bigl(\{r^{(1)},\dots,r^{(N)}\}\bigr)}
                    {\operatorname{std}\!\bigl(\{r^{(1)},\dots,r^{(N)}\}\bigr)} \nonumber\\[6pt]
\rho^{(i)}   &= \frac{\pi_{\bm\theta}\!\bigl(\bm o^{(i)}\mid\bm q\bigr)}
                    {\pi_{\text{old}}\!\bigl(\bm o^{(i)}\mid\bm q\bigr)} \\[6pt]
\mathbb{D}_{\text{KL}}
&= \frac{\pi_{\text{ref}}\!\bigl(\bm o_i\mid\bm q\bigr)}{\pi_{\bm\theta}\!\bigl(\bm o_i\mid\bm q\bigr)}
   - \log\frac{\pi_{\text{ref}}\!\bigl(\bm o_i\mid\bm q\bigr)}{\pi_{\bm\theta}\!\bigl(\bm o_i\mid\bm q\bigr)} - 1 .\nonumber
\end{align}

As the cornerstone of our reinforcement learning, we leverage a mixed-reward mechanism that differs from DeepSeek-R1-like works. 
Contrary to the binary (0/1) reward design which specially evaluate mathematical reasoning task performance based on the final answer correctness, our visual ad summarization task adopts human-annotated reference summaries as the supervision signals. 
Specifically, with the ground truth summaries, we employ lexical coverage and semantic congruence assessment to encourage or punish the policy's generating responses.

\paragraph{\textbf{Lexical Reward (RewardLex)}} In neural machine translation, metrics like BLEU~\cite{BLEU} are typically used to evaluate the quality of model outputs by referencing human-annotated reference translations as the ground truth.
Following this way, we also take BLEU as our lexical reward score function, for encouraging the model's generation to cover key advertising elements, including product subjects, merchant brands, and selling points.

\paragraph{\textbf{Semantic Reward (RewardSem)}} To maintain the semantic integrity of golden (human-annotated) ad video summaries while allowing for paraphrasing, we utilize LLM as a judge. 
Specifically, we provide the LLM (Doubao) with both the human-annotated reference summary and the model-generated summary, prompting it to assess whether the latter sufficiently covers all key points from the reference. 
Besides, we also focus on generating concise summaries to enhance clarity and usability. 
The prompt we use is similar to Figure~\ref{fig:prompt_quality_verification} but is additionally provided a human-written summary for reference.

\paragraph{\textbf{Length Penalty}}
Moreover, to achieve low-latency requirements for online services and prevent the generated summaries from becoming excessively lengthy during reinforcement learning training, we introduce a penalty term in the semantic reward to constrain the length of generated summaries, as formalized in the following equation:
\begin{equation}
\text{LP}(\bm o, \bm y) = \min\left(1, \frac{|\bm y|}{|\bm o|}\right)^\gamma,
\end{equation}
where LP means ``Length Penalty'', $|\bm y|$ is the length of human-written reference summary $\bm y$, $|\bm o|$ denotes the length of model rollout response $\bm o$, and $\gamma\geq 0$ is a hyperparameter controlling the severity of the imposed penalty.

Bringing together all the aforementioned parts, our final mixed-reward function is:
\begin{equation}
\text{R}(\bm o, \bm y) = \text{BLEU}(\bm o, \bm y) + \text{LP}(\bm o, \bm y) \cdot \text{LLM\_Score}(\bm o, \bm y)
\end{equation}

Through this mixed-reward mechanism, we can leverage the verifiably correct references to encourage the policy to generate high-quality outputs - particularly advertising content summaries in our scenario - while simultaneously incentivizing the emergence of semantically similar yet more concise summarization patterns.

\subsection{Online Serving}
\label{section:downstream_tasks}

\begin{figure}[htbp]
  \centering
  \includegraphics[width=0.49\textwidth]{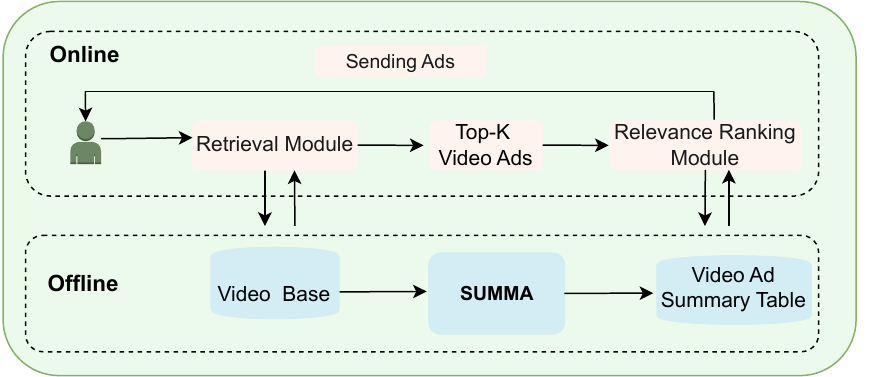}
  \caption{Our online service pipeline employs the SUMMA model to perform inference on all video ads, generating corresponding ad summaries. To ensure real-time performance, these generated summaries are statically stored. Downstream tasks including retrieval and relevance ranking directly utilize the stored summaries of relevant videos for computation. Ultimately, the system delivers ads with the highest relevance scores to the end users.}
  \label{fig:online_serving}
\end{figure}

Given the stringent inference latency requirements in search advertising systems, downstream services typically employ lightweight models with limited parameters. 
As is shown in Figure~\ref{fig:online_serving}, we address this via SUMMA, generating textual summaries that are statically cached in Redis, and thus downstream models can directly fetch these precomputed summaries for inference tasks. 
This approach effectively alleviates the multimodal modeling burden on downstream modules while maintaining low inference latency. By decoupling multimodal processing from task-specific learning, the downstream modules can fully concentrate on their dedicated objectives, thereby achieving enhanced performance. 
Two representative search advertising downstream tasks are as follows:

\noindent (1) The \textbf{Retrieval} stage constitutes the initial phase in search advertising systems, tasked with filtering candidate ad sets from massive advertising inventories. These selected candidates are subsequently forwarded to downstream tasks, including relevance computation and click rate prediction. Our SUMMA framework couples with recall modules by directly utilizing SUMMA-RL inference results, i.e., summaries, as advertisement-side features, which engage in similarity calculations with user-side search queries.

\noindent (2) As for the \textbf{Relevance Ranking} stage, it receives coarsely-ranked advertisement candidates from upstream filtration processes, and computes relevance scores between user queries and advertisements. thereafter, it filters out irrelevant cases. To enhance the precision of relevance computation, we consistently employ the output features generated by SUMMA-RL as critical advertising attributes.

\section{Experiments}

In this section, we first delineate the evaluation metrics and experimental configurations that form the foundation of our investigation. 
Subsequently, comprehensive analyses of both controlled-environment (offline) experiments and real-world deployment (online) trials are provided, accompanied by their respective empirical outcomes.

\subsection{Evaluation Metrics} 
\label{section:evaluation_metrics} 
We employ the following metrics for offline and online evaluations.

\paragraph{\textbf{Offline Metrics}} The offline evaluation primarily aims to provide a reliable and multi-faceted benchmark by measuring summary quality, relevant-ad recall, and fine-grained ranking discrimination, thereby establishing a solid basis for subsequent online experiments.
\begin{itemize}
    \item \textit{BLEU}: N\,-gram precision metric evaluating how closely the generated summary matches the reference text.
    \item \textit{ROUGE}: N\,-gram or longest-common-subsequence recall indicating how much of the reference content is covered by the summary.
    \item \textit{CIDEr}: TF–IDF-weighted n-gram similarity capturing consensus between the summary and multiple references.
    \item \textit{RewardSem}: Cosine similarity of sentence embeddings quantifying semantic closeness between generated and reference texts.
    \item \textit{Hit@\textit{k}}: For each query, the fraction of ground-truth ads appearing in the top-\textit{k} retrieved results.
    \item \textit{ROC–AUC}: Area under the ROC curve representing the probability that a random positive is ranked above a random negative.
    \item \textit{Recall@Precision90}: Recall attained while keeping precision at or above the fixed precision 90\,\%.
\end{itemize}

\paragraph{\textbf{Online Metrics}}

\noindent We conduct online A/B testing to compare SUMMA with the baseline models. Three metrics are considered, as follows:

\begin{itemize}
    \item \textit{Irrelevant Ratio}: The proportion of cases within a sampled dataset that are labeled as ``Bad'' by human annotators, calculated as $\frac{\#\text{Bad} }{\#\text{Bad} +  \#\text{Good} }$.
    
    \item \textit{Conversion Ratio}: An advertising conversion occurs when a potential customer views a video ad and subsequently takes an action deemed valuable to the advertiser's business, such as making an online purchase or calling the business from a mobile phone. We define Conversion Ratio as $\frac{\#\text{Conversion} }{\#\text{Click}}$.
    
    \item \textit{Ad Revenue}: The revenue of the advertising system after distinct methods engaging and assisting advertisers in achieving conversions.
\end{itemize}

\subsection{Experiment Setup} 
The number of frames extracted from a short video is set to 16, which is determined by our preliminary experiments to strike an effective balance between performance and efficiency. During both the SFT and RL stages, we conduct training on one machine comprising 8 NVIDIA A100 GPUs (80GB). The training framework utilized is verl~\cite{verl} based on PyTorch. 

For SFT, we take Qwen2-VL as our base model to train SUMMA-SFT, with a learning rate of 5e-7. For learning rate scheduler, we adopt a cosine decay strategy and a linear warm-up of 2000 steps. We use the AdamW optimizer with $\beta_1$= 0.9, $\beta_2$= 0.999, and a weight decay rate of 0.02.

For GRPO, SUMMA-SFT is used to initialized the policy. The hyperparameters are configured as follows: 8 rollouts are sampled per prompt, while PPO optimization utilizes a batch size of 64. The actor learning rate is 1e-6, and we employ KL coefficient $\beta=0.001$ to stabilize the training process.

\subsection{Offline Experiments} 

Our offline experiments are designed to address the following research questions (RQ):

\noindent\textbf{RQ1.}\; In the context of search advertising summarization task, how to make full use of the multimodal ad content to optimize the MLLMs for generating semantically and commercially impactful ad summaries?

\noindent\textbf{RQ2.}\; How much performance gain can multimodal information bring compared to unimodal approaches (only ocr and asr or frames)? 
    
\noindent\textbf{RQ3.}\; The impact of the richness of OCR/ASR information?
    
\noindent\textbf{RQ4.}\; How do different reward designs affect model performance?
    
\noindent\textbf{RQ5.}\; Compared to the multimodal embedding methods, what improvements can our approach bring to downstream tasks?

\paragraph{\textbf{SFT → RL is the most effective training pipeline (RQ1)}}
To evaluate the summary quality generated from our SUMMA-RL, we compare it with some baseline models on AdSum-Test (see Section~\ref{section:data_construction}) across four key metrics: BLEU, CIDEr, ROUGE, and RSemantic (see Section~\ref{section:evaluation_metrics}). All involved models participating in the comparison consist of:
\begin{itemize}
    \item Qwen2-VL different variants (2B and 7B Instruct versions)
    \item Our SUMMA-SFT model, which is only trained via SFT on AdSum-Doubao
    \item A separate \textbf{two-stage SFT model}, which has undergone SFT on AdSum-Doubao and then SFT on AdSum-Human
    \item A merging \textbf{single-stage SFT model}, directly fine-tuned on the combined dataset of AdSum-Doubao and AdSum-Human
    \item Our SUMMA-RL is derived from SFT on AdSum-Doubao, followed by RL on AdSum-Human.
\end{itemize}

\begin{figure}[htbp]
  \centering
  \includegraphics[width=0.49\textwidth]{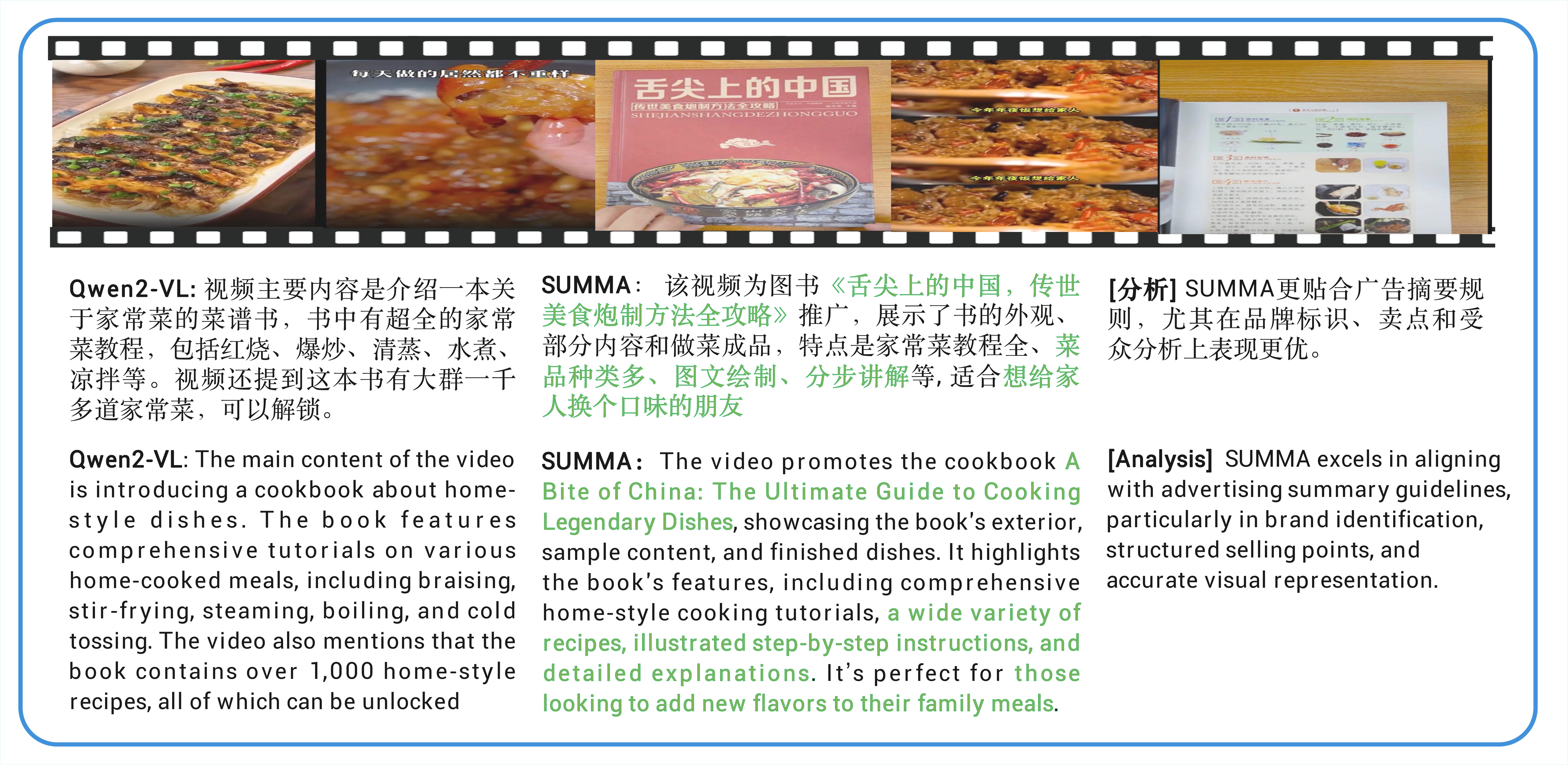}
  \caption{Case study of video ad summarization. SUMMA clearly points out the book's name, characteristics, selling points as well as the target audience, showing the superiority of SUMMA over the base model.}
  \label{fig:case_study}
\end{figure}

\begin{table}[htbp]
\centering
\caption{The performances of different models on the AdSum-Test across various evaluation metrics. The best result for each metric is boldfaced.}
\label{tab:model_performance}
\resizebox{.48\textwidth}{!}{
\begin{tabular}{lcccc}
\toprule
\textbf{Method} & \textbf{BLEU} &\textbf{CIDEr} & \textbf{ROUGE} & \textbf{RewardSem} \\
\midrule
Qwen2-VL-2B-Instruct & 0.10 & 0.156 & 0.26 & 0.09 \\
Qwen2-VL-7B-Instruct & 0.14 & 0.18 & 0.29 & 0.12 \\
SUMMA-SFT & 0.36 & 1.01 & 0.46 & 0.26 \\
\shortstack{Single-Stage SFT} & 0.38 & 1.05 & 0.48 & 0.28 \\
\shortstack{Two-Stage SFT} & 0.41 & 1.12 & 0.52 & 0.30 \\
\shortstack{SUMMA-RL} & \textbf{0.48} & \textbf{1.42} & \textbf{0.62} & \textbf{0.38} \\
\bottomrule
\end{tabular}
}
\end{table}
As shown in Table~\ref{tab:model_performance}, our method demonstrates consistent improvements across all evaluation metrics. 
Even our intermediary SUMMA-SFT model can outperform both Qwen2-VL models by significant margins (e.g., 0.36 vs 0.14 in BLEU, and 1.01 vs 0.18 in CIDEr). 
This performance gap highlights the effectiveness of our constructed datasets, which infuse domain-specific knowledge into the model and thus enhance its capability in search advertising scenarios. 
Moreover, compared to the other training strategies (two-stage SFT or mixed 1-stage SFT), ``SFT then RL'' is found to be the best one, providing additional evidence that the training paradigm also remains effective when applied to the search advertising domain, and thus demonstrating its broader applicability. 
Figure~\ref{fig:case_study} is a case study showing the superiority of the summary generated from SUMMA over the base model.

\begin{table}[htbp]
\caption{The synergistic effects between different modalities. Multimodal (video + OCR/ASR) input outperforms either single modality}
\centering
\setlength{\tabcolsep}{3pt}
\label{tab:modal_impact}
\begin{tabular}{lcccc}
\toprule
\textbf{Method} & \textbf{BLEU} & \textbf{CIDEr} & \textbf{ROUGE} & \textbf{RewardSem} \\
\midrule
Unimodal-OCR/ASR & 0.27 & 0.64 & 0.42 & 0.17 \\
Unimodal-Video & 0.34 & 0.82 & 0.48 & 0.21 \\
Multimodal & \textbf{0.48} & \textbf{1.42} & \textbf{0.62} & \textbf{0.38} \\
\bottomrule
\end{tabular}
\end{table}

\paragraph{\textbf{Multimodal input clearly outperforms any single modality (RQ2)}}
To evaluate the efficacy of multimodal integration, we conduct a controlled ablation study comparing three different training configurations: unimodal (video-only), unimodal (OCR/ASR-only) and multimodal (video+OCR/ASR). As evidenced in Table~\ref{tab:modal_impact}, combining different modalities, our multimodal approach demonstrates statistically significant improvements across all evaluation metrics. 
Specifically, compared to the best unimodal baseline, we observe relative improvements of +0.14 (BLEU), +0.60 (CIDEr), +0.14 (ROUGE), and +0.17 (RewardSem), underscoring the complementary nature of visual and textual modalities. 
Therefore, the integration of linguistic signals from ASR/OCR with visual features can provide significant advantages for comprehensive video understanding and summarization by capturing rich semantic context than any single modality alone.

\paragraph{\textbf{More OCR/ASR text brings larger gains (RQ3)}}
We collect two other distinct datasets, each comprising 2,500 samples, categorized by the richness of OCR/ASR textual information. This stratification is designed to evaluate how OCR/ASR information density affects the same SUMMA-RL model. The ``Sparse'' category denotes samples with combined OCR/ASR character counts \textbf{below 10}, while ``Rich'' represents the converse condition. Table~\ref{tab:ocr_ablation} shows that the model performs better as the OCR information content increases.
\begin{table}[htbp]
\centering
\caption{Performance of SUMMA-RL on the AdSum-Test set, which we split into two subsets according to ASR/OCR information density: \textit{Sparse} vs.\ \textit{Rich}.}
\label{tab:ocr_ablation}
\resizebox{.47\textwidth}{!}{
\begin{tabular}{lcccc}
\toprule
\textbf{Test Set} & \textbf{BLEU} & \textbf{CIDEr} & \textbf{ROUGE} & \textbf{RewardSem} \\
\midrule
Sparse ASR/OCR & 0.28 & 0.65 & 0.46 & 0.19 \\
Rich ASR/OCR & \textbf{0.52} & \textbf{1.53} & \textbf{0.69} & \textbf{0.45} \\
AdSum-Test & 0.48 & 1.42 & 0.62 & 0.38 \\
\bottomrule
\end{tabular}
}
\end{table}

\paragraph{\textbf{Mixed lexical + semantic rewards beat single-aspect rewards (RQ4)}}

We conduct ablation studies to investigate the effectiveness of our mixed-reward mechanism. 
As demonstrated in Table~\ref{tab:reward_design}, models trained with the mixed reward achieve obviously better performance compared to those trained solely with either lexical or semantic reward. 
This empirical evidence suggests that rewards from these two distinct perspectives exhibit synergistic complementarity, enabling the final combined reward to effectively incentivize the model to generate high-quality ad summaries.

\begin{table}[htbp]
\centering
\caption{Ablation study of reward design: lexical-only, semantic-only, and mixed (RewardLex + RewardSem).}
\label{tab:reward_design}
\begin{tabular}{lcccc}
\toprule
\textbf{Method} & \textbf{BLEU} & \textbf{CIDEr} & \textbf{ROUGE} & \textbf{RewardSem} \\
\midrule
\shortstack{RewardLex} & 0.44 & 1.24 & 0.57 & 0.31 \\
\shortstack{RewardSem} & 0.38 & 1.11 & 0.49 & 0.37 \\
\shortstack{RewardLex \\+ RewardSem} & \textbf{0.48} & \textbf{1.42} & \textbf{0.62} & \textbf{0.38} \\
\bottomrule
\end{tabular}
\end{table}

\begin{figure}[htbp]
  \centering
  \includegraphics[width=0.48\textwidth]{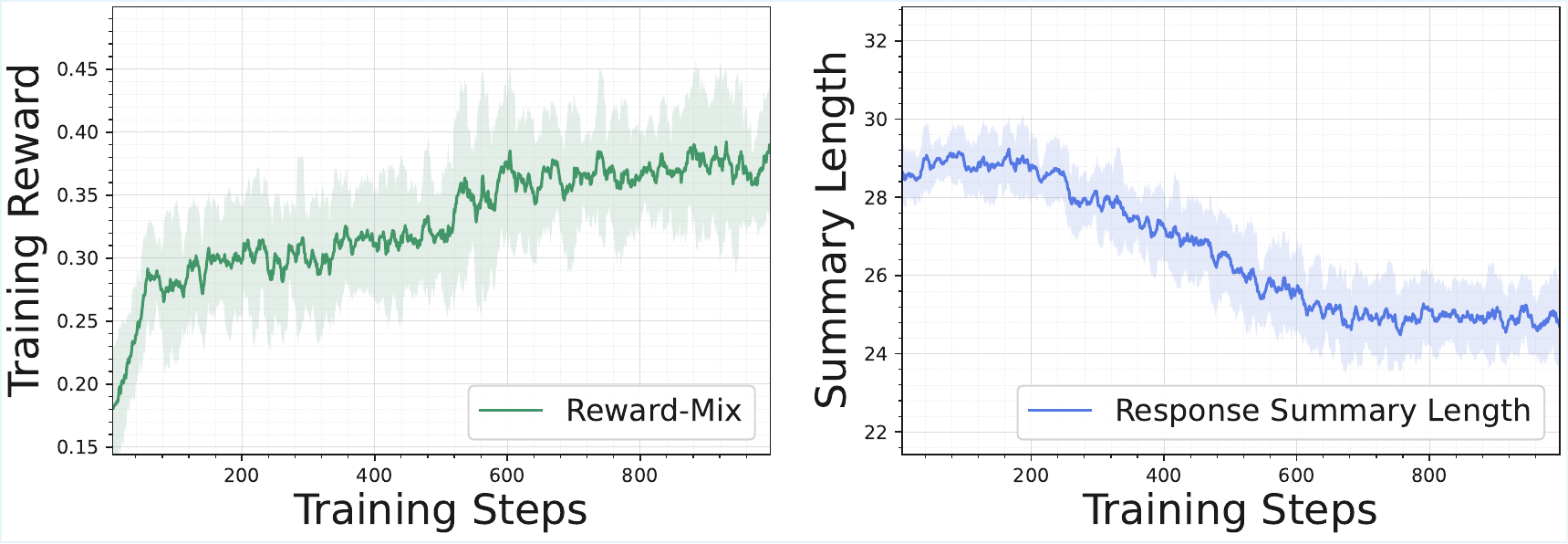}
  \caption{The progressive changes in training rewards and response lengths with respect to the global steps during the RL process.}
  \label{fig:training_reward}
\end{figure}

Figure~\ref{fig:training_reward} illustrates the evolution of training rewards and generated response lengths throughout the training process. 
As shown, the reward values exhibit a consistent upward trend during the initial phase, eventually converging to a stable plateau with minimal oscillations. Concurrently, the summary length remains relatively stable in the early stages, followed by a modest increase and subsequent gradual decline. 
This trajectory clearly demonstrates the effectiveness of our designed reward mechanism.

\paragraph{\textbf{High-quality summaries enhance downstream retrieval and ranking (RQ5)}}

For diverse downstream tasks (i.e., Retrieval and Relevance Ranking, see Section~\ref{section:downstream_tasks}), we compare SUMMA against multimodal embedding models. 
In retrieval, our in-house evaluation dataset consists of 30K query-ad pairs and 5M candidate ads, while the relevance ranking test set consists of 5K query-ad pairs.

As shown in Table~\ref{tab:retrieval_comparison} and Table~\ref{tab:relevance_comparison}, SUMMA demonstrates significant superiority over embedding-based methods in both search advertising tasks. 
These empirical results substantiate that SUMMA achieves more adequate utilization and sophisticated modeling of multimodal advertising contents, thereby validating the efficacy of our research trajectory.

Additionally, to evaluate the advantages of our method, we conduct assessments from the perspectives of granularity and diversity.
 First, we randomly select 10K queries, perform ANN searches to retrieve their top-100 similar ads from each index, and calculate the proportion of unique count of retrieved ads to all ads as the \textbf{diversity ratio}. 
Second, we define granularity ratio to quantify the impact of granularity perspectives on retrieval performance. Specifically, as the granularity of ad features becomes finer, the diversity of retrieved results for similar queries should correspondingly improve. Therefore, we select 10K queries with minimal semantic differences and repeat the same retrieval process as computing diversity ratio. As shown in Table~\ref{tab:retrieval_comparison},
the diversity ratio increases by over 3\% and the granularity ratio increases by over 2\%, indicating our SUMMA can expand the diversity of retrieved advertisements while simultaneously excels in achieving granular discrimination among ad candidates.
 
\begin{table}[htbp]
\centering
\caption{Performance comparison between our SUMMA and the embedding method for the downstream retrieval task.}
\label{tab:retrieval_comparison}
\setlength{\tabcolsep}{4pt} 
\begin{tabular}{lccccc}
\toprule
\textbf{Method} & \makecell[c]{\textbf{Diversity}\\ \textbf{Ratio}} & \makecell[c]{\textbf{Granularity}\\ \textbf{Ratio}} &\textbf{Hit@10} & \textbf{Hit@100}   \\
\midrule
\makecell[l]{Multimodal \\ \;\; Embedding} & 7.22\% & 1.22\% & 4.6\% & 18\% \\
SUMMA & \textbf{10.23\%} & \textbf{3.22\%} & \textbf{6.5\%} & \textbf{22\%}  \\
\bottomrule
\end{tabular}
\end{table}

\begin{table}[htbp]
\centering
\caption{Performance comparison between our SUMMA and the embedding method for the downstream relevance ranking task.}
\label{tab:relevance_comparison}
\setlength{\tabcolsep}{4pt} 
\begin{tabular}{lccccc}
\toprule
\textbf{Method} & \textbf{AUC} &  \shortstack{\textbf{Rec.@Prec.90}}  \\
\midrule
Multimodal Embedding & 0.90 & 0.65  \\
SUMMA & \textbf{0.96} & \textbf{0.71}  \\
\bottomrule
\end{tabular}
\end{table}

\subsection{Online Experiments}
A production-level A/B testing is implemented by incorporating our SUMMA into the BERT-based retrieval and relevance models within our real-time search advertising system. 
While maintaining all other factors unchanged, we expose the experimental variant to 10\% of total search ad traffic of our platform for 14 consecutive days to ensure statistical significance.

As is shown in Table~\ref{tab:online_comparison}, the experimental results demonstrate that our SUMMA yields a statistically significant 1.5\% enhancement in advertising revenue. 
Additionally, through rigorous human evaluation protocols, we observe a 5\% reduction in relevance bad rate. 
Overall, these quantitative improvements collectively validate SUMMA's capacity to optimize the whole search advertising system while simultaneously improving the user experience.

Therefore, our SUMMA has been deployed as the core component supporting our entire search advertising ecosystem. 
Moreover, we note that beyond its fundamental role in powering retrieval and relevance ranking, SUMMA can also be utilized in enhancing feature integration for other ranking models like the ones targeting Click-Through Rate (CTR) and CVR (Conversion Rate) prediction.

In conclusion, the practical deployment of our SUMMA on the real-world online search ad system demonstrates its measurable benefits in production environments.

\begin{table}[htbp]
\centering
\caption{Online A/B-test gains of SUMMA: irrelevant-case ratio (↓), conversion ratio (↑), and ad revenue (↑).}
\label{tab:online_comparison}
\resizebox{.48\textwidth}{!}{
\begin{tabular}{lcccc}
\toprule
\textbf{Task} & \makecell[c]{\textbf{Irrelevant Ratio}\\ \textbf{$\bm \Delta$ ($\bm \downarrow$)}} & \makecell[c]{\textbf{Conversion Ratio}\\ \textbf{$\bm \Delta$ ($\bm \uparrow$)}} & \makecell[c]{\textbf{Ad Revenue}\\ \textbf{$\bm \Delta$ ($\bm \uparrow$)}} \\
\midrule
\textbf{Retrieval} & -1.0\% & +0.9\% & +0.5\% \\
\textbf{Relevance Ranking} & -4.0\% & +1.5\% & +1.0\% \\
\bottomrule
\end{tabular}
}
\end{table}

\section{Conclusion}
In this paper, we present SUMMA, a novel MLLM trained on our newly constructed multimodal advertising summarization datasets for supervised fine-tuning and reinforcement learning. 
As the first OCR/ASR-coordinated multimodal cognitive approach for search advertising, SUMMA can effectively process visual-text features in advertising videos to generate commercially valuable and concise video summaries.
Besides, we further implement a pipeline integrated in the search advertising system where our SUMMA serves as the upstream component and synergizes with various downstream modules to perform candidate retrieval and relevance ranking. 
Empirical validation through offline experiments and online production environment testing confirms SUMMA's superior performance.
In our future research, we will explore more effective MLLMs RL strategies, such as incorporating more downstream tasks into the reward design to further refine the quality of multimodal advertising summaries, ultimately leading to more accurate user-ad matching and more enhanced user experience.

\section*{GenAI Usage Disclosure}
We use Doubao to generate and verify our ad summarization SFT Data (AdSum-Doubao), and also employ it as the judge model to determine the semantic rewards in our RL training stage.


\bibliographystyle{ACM-Reference-Format}
\bibliography{sample-base}


\begin{thebibliography}{74}


\ifx \showCODEN    \undefined \def \showCODEN     #1{\unskip}     \fi
\ifx \showISBNx    \undefined \def \showISBNx     #1{\unskip}     \fi
\ifx \showISBNxiii \undefined \def \showISBNxiii  #1{\unskip}     \fi
\ifx \showISSN     \undefined \def \showISSN      #1{\unskip}     \fi
\ifx \showLCCN     \undefined \def \showLCCN      #1{\unskip}     \fi
\ifx \shownote     \undefined \def \shownote      #1{#1}          \fi
\ifx \showarticletitle \undefined \def \showarticletitle #1{#1}   \fi
\ifx \showURL      \undefined \def \showURL       {\relax}        \fi
\providecommand\bibfield[2]{#2}
\providecommand\bibinfo[2]{#2}
\providecommand\natexlab[1]{#1}
\providecommand\showeprint[2][]{arXiv:#2}

\bibitem[AI(2024)]%
        {Llama-3}
\bibfield{author}{\bibinfo{person}{Meta AI}.} \bibinfo{year}{2024}\natexlab{}.
\newblock \bibinfo{title}{The Llama 3 Herd of Models}.
\newblock
\showeprint[arxiv]{2407.21783}~[cs.AI]
\urldef\tempurl%
\url{https://arxiv.org/abs/2407.21783}
\showURL{%
\tempurl}


\bibitem[Alayrac et~al\mbox{.}(2022)]%
        {Flamingo}
\bibfield{author}{\bibinfo{person}{Jean-Baptiste Alayrac}, \bibinfo{person}{Jeff Donahue}, \bibinfo{person}{Pauline Luc}, \bibinfo{person}{Antoine Miech}, \bibinfo{person}{Iain Barr}, \bibinfo{person}{Yana Hasson}, \bibinfo{person}{Karel Lenc}, \bibinfo{person}{Arthur Mensch}, \bibinfo{person}{Katie Millican}, \bibinfo{person}{Malcolm Reynolds}, \bibinfo{person}{Roman Ring}, \bibinfo{person}{Eliza Rutherford}, \bibinfo{person}{Serkan Cabi}, \bibinfo{person}{Tengda Han}, \bibinfo{person}{Zhitao Gong}, \bibinfo{person}{Sina Samangooei}, \bibinfo{person}{Marianne Monteiro}, \bibinfo{person}{Jacob Menick}, \bibinfo{person}{Sebastian Borgeaud}, \bibinfo{person}{Andrew Brock}, \bibinfo{person}{Aida Nematzadeh}, \bibinfo{person}{Sahand Sharifzadeh}, \bibinfo{person}{Mikolaj Binkowski}, \bibinfo{person}{Ricardo Barreira}, \bibinfo{person}{Oriol Vinyals}, \bibinfo{person}{Andrew Zisserman}, {and} \bibinfo{person}{Karen Simonyan}.} \bibinfo{year}{2022}\natexlab{}.
\newblock \bibinfo{title}{Flamingo: a Visual Language Model for Few-Shot Learning}.
\newblock
\showeprint[arxiv]{2204.14198}~[cs.CV]
\urldef\tempurl%
\url{https://arxiv.org/abs/2204.14198}
\showURL{%
\tempurl}


\bibitem[Bai et~al\mbox{.}(2023)]%
        {Qwen-VL}
\bibfield{author}{\bibinfo{person}{Jinze Bai}, \bibinfo{person}{Shuai Bai}, \bibinfo{person}{Shusheng Yang}, \bibinfo{person}{Shijie Wang}, \bibinfo{person}{Sinan Tan}, \bibinfo{person}{Peng Wang}, \bibinfo{person}{Junyang Lin}, \bibinfo{person}{Chang Zhou}, {and} \bibinfo{person}{Jingren Zhou}.} \bibinfo{year}{2023}\natexlab{}.
\newblock \bibinfo{title}{Qwen-VL: A Versatile Vision-Language Model for Understanding, Localization, Text Reading, and Beyond}.
\newblock
\showeprint[arxiv]{2308.12966}~[cs.CV]
\urldef\tempurl%
\url{https://arxiv.org/abs/2308.12966}
\showURL{%
\tempurl}


\bibitem[Bai et~al\mbox{.}(2025)]%
        {Qwen2.5-VL}
\bibfield{author}{\bibinfo{person}{Shuai Bai}, \bibinfo{person}{Keqin Chen}, \bibinfo{person}{Xuejing Liu}, \bibinfo{person}{Jialin Wang}, \bibinfo{person}{Wenbin Ge}, \bibinfo{person}{Sibo Song}, \bibinfo{person}{Kai Dang}, \bibinfo{person}{Peng Wang}, \bibinfo{person}{Shijie Wang}, \bibinfo{person}{Jun Tang}, \bibinfo{person}{Humen Zhong}, \bibinfo{person}{Yuanzhi Zhu}, \bibinfo{person}{Mingkun Yang}, \bibinfo{person}{Zhaohai Li}, \bibinfo{person}{Jianqiang Wan}, \bibinfo{person}{Pengfei Wang}, \bibinfo{person}{Wei Ding}, \bibinfo{person}{Zheren Fu}, \bibinfo{person}{Yiheng Xu}, \bibinfo{person}{Jiabo Ye}, \bibinfo{person}{Xi Zhang}, \bibinfo{person}{Tianbao Xie}, \bibinfo{person}{Zesen Cheng}, \bibinfo{person}{Hang Zhang}, \bibinfo{person}{Zhibo Yang}, \bibinfo{person}{Haiyang Xu}, {and} \bibinfo{person}{Junyang Lin}.} \bibinfo{year}{2025}\natexlab{}.
\newblock \bibinfo{title}{Qwen2.5-VL Technical Report}.
\newblock
\showeprint[arxiv]{2502.13923}~[cs.CV]
\urldef\tempurl%
\url{https://arxiv.org/abs/2502.13923}
\showURL{%
\tempurl}


\bibitem[ByteDance(2025)]%
        {Doubao-1.5-pro}
\bibfield{author}{\bibinfo{person}{ByteDance}.} \bibinfo{year}{2025}\natexlab{}.
\newblock \bibinfo{title}{Doubao-1.5-pro}.
\newblock
\urldef\tempurl%
\url{https://seed.bytedance.com/en/special/doubao_1_5_pro}
\showURL{%
\tempurl}


\bibitem[Cai et~al\mbox{.}(2024)]%
        {InternLM2}
\bibfield{author}{\bibinfo{person}{Zheng Cai}, \bibinfo{person}{Maosong Cao}, \bibinfo{person}{Haojiong Chen}, \bibinfo{person}{Kai Chen}, \bibinfo{person}{Keyu Chen}, \bibinfo{person}{Xin Chen}, \bibinfo{person}{Xun Chen}, \bibinfo{person}{Zehui Chen}, \bibinfo{person}{Zhi Chen}, \bibinfo{person}{Pei Chu}, \bibinfo{person}{Xiaoyi Dong}, \bibinfo{person}{Haodong Duan}, \bibinfo{person}{Qi Fan}, \bibinfo{person}{Zhaoye Fei}, \bibinfo{person}{Yang Gao}, \bibinfo{person}{Jiaye Ge}, \bibinfo{person}{Chenya Gu}, \bibinfo{person}{Yuzhe Gu}, \bibinfo{person}{Tao Gui}, \bibinfo{person}{Aijia Guo}, \bibinfo{person}{Qipeng Guo}, \bibinfo{person}{Conghui He}, \bibinfo{person}{Yingfan Hu}, \bibinfo{person}{Ting Huang}, \bibinfo{person}{Tao Jiang}, \bibinfo{person}{Penglong Jiao}, \bibinfo{person}{Zhenjiang Jin}, \bibinfo{person}{Zhikai Lei}, \bibinfo{person}{Jiaxing Li}, \bibinfo{person}{Jingwen Li}, \bibinfo{person}{Linyang Li}, \bibinfo{person}{Shuaibin Li}, \bibinfo{person}{Wei Li}, \bibinfo{person}{Yining Li},
  \bibinfo{person}{Hongwei Liu}, \bibinfo{person}{Jiangning Liu}, \bibinfo{person}{Jiawei Hong}, \bibinfo{person}{Kaiwen Liu}, \bibinfo{person}{Kuikun Liu}, \bibinfo{person}{Xiaoran Liu}, \bibinfo{person}{Chengqi Lv}, \bibinfo{person}{Haijun Lv}, \bibinfo{person}{Kai Lv}, \bibinfo{person}{Li Ma}, \bibinfo{person}{Runyuan Ma}, \bibinfo{person}{Zerun Ma}, \bibinfo{person}{Wenchang Ning}, \bibinfo{person}{Linke Ouyang}, \bibinfo{person}{Jiantao Qiu}, \bibinfo{person}{Yuan Qu}, \bibinfo{person}{Fukai Shang}, \bibinfo{person}{Yunfan Shao}, \bibinfo{person}{Demin Song}, \bibinfo{person}{Zifan Song}, \bibinfo{person}{Zhihao Sui}, \bibinfo{person}{Peng Sun}, \bibinfo{person}{Yu Sun}, \bibinfo{person}{Huanze Tang}, \bibinfo{person}{Bin Wang}, \bibinfo{person}{Guoteng Wang}, \bibinfo{person}{Jiaqi Wang}, \bibinfo{person}{Jiayu Wang}, \bibinfo{person}{Rui Wang}, \bibinfo{person}{Yudong Wang}, \bibinfo{person}{Ziyi Wang}, \bibinfo{person}{Xingjian Wei}, \bibinfo{person}{Qizhen Weng}, \bibinfo{person}{Fan Wu},
  \bibinfo{person}{Yingtong Xiong}, \bibinfo{person}{Chao Xu}, \bibinfo{person}{Ruiliang Xu}, \bibinfo{person}{Hang Yan}, \bibinfo{person}{Yirong Yan}, \bibinfo{person}{Xiaogui Yang}, \bibinfo{person}{Haochen Ye}, \bibinfo{person}{Huaiyuan Ying}, \bibinfo{person}{Jia Yu}, \bibinfo{person}{Jing Yu}, \bibinfo{person}{Yuhang Zang}, \bibinfo{person}{Chuyu Zhang}, \bibinfo{person}{Li Zhang}, \bibinfo{person}{Pan Zhang}, \bibinfo{person}{Peng Zhang}, \bibinfo{person}{Ruijie Zhang}, \bibinfo{person}{Shuo Zhang}, \bibinfo{person}{Songyang Zhang}, \bibinfo{person}{Wenjian Zhang}, \bibinfo{person}{Wenwei Zhang}, \bibinfo{person}{Xingcheng Zhang}, \bibinfo{person}{Xinyue Zhang}, \bibinfo{person}{Hui Zhao}, \bibinfo{person}{Qian Zhao}, \bibinfo{person}{Xiaomeng Zhao}, \bibinfo{person}{Fengzhe Zhou}, \bibinfo{person}{Zaida Zhou}, \bibinfo{person}{Jingming Zhuo}, \bibinfo{person}{Yicheng Zou}, \bibinfo{person}{Xipeng Qiu}, \bibinfo{person}{Yu Qiao}, {and} \bibinfo{person}{Dahua Lin}.} \bibinfo{year}{2024}\natexlab{}.
\newblock \bibinfo{title}{InternLM2 Technical Report}.
\newblock
\showeprint[arxiv]{2403.17297}~[cs.CL]
\urldef\tempurl%
\url{https://arxiv.org/abs/2403.17297}
\showURL{%
\tempurl}


\bibitem[Chang et~al\mbox{.}(2021)]%
        {XMC}
\bibfield{author}{\bibinfo{person}{Wei-Cheng Chang}, \bibinfo{person}{Daniel Jiang}, \bibinfo{person}{Hsiang-Fu Yu}, \bibinfo{person}{Choon-Hui Teo}, \bibinfo{person}{Jiong Zhang}, \bibinfo{person}{Kai Zhong}, \bibinfo{person}{Kedarnath Kolluri}, \bibinfo{person}{Qie Hu}, \bibinfo{person}{Nikhil Shandilya}, \bibinfo{person}{Vyacheslav Ievgrafov}, \bibinfo{person}{Japinder Singh}, {and} \bibinfo{person}{Inderjit~S. Dhillon}.} \bibinfo{year}{2021}\natexlab{}.
\newblock \bibinfo{title}{Extreme Multi-label Learning for Semantic Matching in Product Search}.
\newblock
\showeprint[arxiv]{2106.12657}~[cs.IR]
\urldef\tempurl%
\url{https://arxiv.org/abs/2106.12657}
\showURL{%
\tempurl}


\bibitem[Chang et~al\mbox{.}(2025)]%
        {BLEUBERI}
\bibfield{author}{\bibinfo{person}{Yapei Chang}, \bibinfo{person}{Yekyung Kim}, \bibinfo{person}{Michael Krumdick}, \bibinfo{person}{Amir Zadeh}, \bibinfo{person}{Chuan Li}, \bibinfo{person}{Chris Tanner}, {and} \bibinfo{person}{Mohit Iyyer}.} \bibinfo{year}{2025}\natexlab{}.
\newblock \bibinfo{title}{BLEUBERI: BLEU is a surprisingly effective reward for instruction following}.
\newblock
\showeprint[arxiv]{2505.11080}~[cs.CL]
\urldef\tempurl%
\url{https://arxiv.org/abs/2505.11080}
\showURL{%
\tempurl}


\bibitem[Chen et~al\mbox{.}(2024a)]%
        {MMStar}
\bibfield{author}{\bibinfo{person}{Lin Chen}, \bibinfo{person}{Jinsong Li}, \bibinfo{person}{Xiaoyi Dong}, \bibinfo{person}{Pan Zhang}, \bibinfo{person}{Yuhang Zang}, \bibinfo{person}{Zehui Chen}, \bibinfo{person}{Haodong Duan}, \bibinfo{person}{Jiaqi Wang}, \bibinfo{person}{Yu Qiao}, \bibinfo{person}{Dahua Lin}, {and} \bibinfo{person}{Feng Zhao}.} \bibinfo{year}{2024}\natexlab{a}.
\newblock \bibinfo{title}{Are We on the Right Way for Evaluating Large Vision-Language Models?}
\newblock
\showeprint[arxiv]{2403.20330}~[cs.CV]
\urldef\tempurl%
\url{https://arxiv.org/abs/2403.20330}
\showURL{%
\tempurl}


\bibitem[Chen et~al\mbox{.}(2024b)]%
        {InternVL2.5}
\bibfield{author}{\bibinfo{person}{Zhe Chen}, \bibinfo{person}{Weiyun Wang}, \bibinfo{person}{Yue Cao}, \bibinfo{person}{Yangzhou Liu}, \bibinfo{person}{Zhangwei Gao}, \bibinfo{person}{Erfei Cui}, \bibinfo{person}{Jinguo Zhu}, \bibinfo{person}{Shenglong Ye}, \bibinfo{person}{Hao Tian}, \bibinfo{person}{Zhaoyang Liu}, {et~al\mbox{.}}} \bibinfo{year}{2024}\natexlab{b}.
\newblock \showarticletitle{Expanding Performance Boundaries of Open-Source Multimodal Models with Model, Data, and Test-Time Scaling}.
\newblock \bibinfo{journal}{\emph{arXiv preprint arXiv:2412.05271}} (\bibinfo{year}{2024}).
\newblock


\bibitem[Chen et~al\mbox{.}(2024c)]%
        {InternVL1.5}
\bibfield{author}{\bibinfo{person}{Zhe Chen}, \bibinfo{person}{Weiyun Wang}, \bibinfo{person}{Hao Tian}, \bibinfo{person}{Shenglong Ye}, \bibinfo{person}{Zhangwei Gao}, \bibinfo{person}{Erfei Cui}, \bibinfo{person}{Wenwen Tong}, \bibinfo{person}{Kongzhi Hu}, \bibinfo{person}{Jiapeng Luo}, \bibinfo{person}{Zheng Ma}, {et~al\mbox{.}}} \bibinfo{year}{2024}\natexlab{c}.
\newblock \showarticletitle{How far are we to gpt-4v? closing the gap to commercial multimodal models with open-source suites}.
\newblock \bibinfo{journal}{\emph{Science China Information Sciences}} \bibinfo{volume}{67}, \bibinfo{number}{12} (\bibinfo{year}{2024}), \bibinfo{pages}{220101}.
\newblock


\bibitem[Chen et~al\mbox{.}(2024d)]%
        {InternVL}
\bibfield{author}{\bibinfo{person}{Zhe Chen}, \bibinfo{person}{Jiannan Wu}, \bibinfo{person}{Wenhai Wang}, \bibinfo{person}{Weijie Su}, \bibinfo{person}{Guo Chen}, \bibinfo{person}{Sen Xing}, \bibinfo{person}{Muyan Zhong}, \bibinfo{person}{Qinglong Zhang}, \bibinfo{person}{Xizhou Zhu}, \bibinfo{person}{Lewei Lu}, {et~al\mbox{.}}} \bibinfo{year}{2024}\natexlab{d}.
\newblock \showarticletitle{Internvl: Scaling up vision foundation models and aligning for generic visual-linguistic tasks}. In \bibinfo{booktitle}{\emph{Proceedings of the IEEE/CVF Conference on Computer Vision and Pattern Recognition}}. \bibinfo{pages}{24185--24198}.
\newblock


\bibitem[Dai et~al\mbox{.}(2023)]%
        {InstructBLIP}
\bibfield{author}{\bibinfo{person}{Wenliang Dai}, \bibinfo{person}{Junnan Li}, \bibinfo{person}{Dongxu Li}, \bibinfo{person}{Anthony Meng~Huat Tiong}, \bibinfo{person}{Junqi Zhao}, \bibinfo{person}{Weisheng Wang}, \bibinfo{person}{Boyang Li}, \bibinfo{person}{Pascale Fung}, {and} \bibinfo{person}{Steven Hoi}.} \bibinfo{year}{2023}\natexlab{}.
\newblock \bibinfo{title}{InstructBLIP: Towards General-purpose Vision-Language Models with Instruction Tuning}.
\newblock
\showeprint[arxiv]{2305.06500}~[cs.CV]
\urldef\tempurl%
\url{https://arxiv.org/abs/2305.06500}
\showURL{%
\tempurl}


\bibitem[DeepSeek-AI(2024)]%
        {DeepSeek-V2}
\bibfield{author}{\bibinfo{person}{DeepSeek-AI}.} \bibinfo{year}{2024}\natexlab{}.
\newblock \bibinfo{title}{DeepSeek-V2: A Strong, Economical, and Efficient Mixture-of-Experts Language Model}.
\newblock
\showeprint[arxiv]{2405.04434}~[cs.CL]
\urldef\tempurl%
\url{https://arxiv.org/abs/2405.04434}
\showURL{%
\tempurl}


\bibitem[DeepSeek-AI(2025)]%
        {DeepSeek-R1}
\bibfield{author}{\bibinfo{person}{DeepSeek-AI}.} \bibinfo{year}{2025}\natexlab{}.
\newblock \bibinfo{title}{DeepSeek-R1: Incentivizing Reasoning Capability in LLMs via Reinforcement Learning}.
\newblock
\showeprint[arxiv]{2501.12948}~[cs.CL]
\urldef\tempurl%
\url{https://arxiv.org/abs/2501.12948}
\showURL{%
\tempurl}


\bibitem[Dong et~al\mbox{.}(2024)]%
        {M2-RAAP}
\bibfield{author}{\bibinfo{person}{Xingning Dong}, \bibinfo{person}{Zipeng Feng}, \bibinfo{person}{Chunluan Zhou}, \bibinfo{person}{Xuzheng Yu}, \bibinfo{person}{Ming Yang}, {and} \bibinfo{person}{Qingpei Guo}.} \bibinfo{year}{2024}\natexlab{}.
\newblock \bibinfo{title}{M2-RAAP: A Multi-Modal Recipe for Advancing Adaptation-based Pre-training towards Effective and Efficient Zero-shot Video-text Retrieval}.
\newblock
\showeprint[arxiv]{2401.17797}~[cs.CV]
\urldef\tempurl%
\url{https://arxiv.org/abs/2401.17797}
\showURL{%
\tempurl}


\bibitem[Feng et~al\mbox{.}(2025b)]%
        {Video-R1}
\bibfield{author}{\bibinfo{person}{Kaituo Feng}, \bibinfo{person}{Kaixiong Gong}, \bibinfo{person}{Bohao Li}, \bibinfo{person}{Zonghao Guo}, \bibinfo{person}{Yibing Wang}, \bibinfo{person}{Tianshuo Peng}, \bibinfo{person}{Benyou Wang}, {and} \bibinfo{person}{Xiangyu Yue}.} \bibinfo{year}{2025}\natexlab{b}.
\newblock \bibinfo{title}{Video-R1: Reinforcing Video Reasoning in MLLMs}.
\newblock
\showeprint[arxiv]{2503.21776}~[cs.CV]
\urldef\tempurl%
\url{https://arxiv.org/abs/2503.21776}
\showURL{%
\tempurl}


\bibitem[Feng et~al\mbox{.}(2025a)]%
        {MT-R1-Zero}
\bibfield{author}{\bibinfo{person}{Zhaopeng Feng}, \bibinfo{person}{Shaosheng Cao}, \bibinfo{person}{Jiahan Ren}, \bibinfo{person}{Jiayuan Su}, \bibinfo{person}{Ruizhe Chen}, \bibinfo{person}{Yan Zhang}, \bibinfo{person}{Zhe Xu}, \bibinfo{person}{Yao Hu}, \bibinfo{person}{Jian Wu}, {and} \bibinfo{person}{Zuozhu Liu}.} \bibinfo{year}{2025}\natexlab{a}.
\newblock \bibinfo{title}{MT-R1-Zero: Advancing LLM-based Machine Translation via R1-Zero-like Reinforcement Learning}.
\newblock
\showeprint[arxiv]{2504.10160}~[cs.CL]
\urldef\tempurl%
\url{https://arxiv.org/abs/2504.10160}
\showURL{%
\tempurl}


\bibitem[Gan et~al\mbox{.}(2025)]%
        {HCMRM}
\bibfield{author}{\bibinfo{person}{Guobing Gan}, \bibinfo{person}{Kaiming Gao}, \bibinfo{person}{Li Wang}, \bibinfo{person}{Shen Jiang}, {and} \bibinfo{person}{Peng Jiang}.} \bibinfo{year}{2025}\natexlab{}.
\newblock \bibinfo{title}{HCMRM: A High-Consistency Multimodal Relevance Model for Search Ads}.
\newblock
\showeprint[arxiv]{2502.05822}~[cs.IR]
\urldef\tempurl%
\url{https://arxiv.org/abs/2502.05822}
\showURL{%
\tempurl}


\bibitem[GLM et~al\mbox{.}(2024)]%
        {ChatGLM}
\bibfield{author}{\bibinfo{person}{Team GLM}, \bibinfo{person}{:}, \bibinfo{person}{Aohan Zeng}, \bibinfo{person}{Bin Xu}, \bibinfo{person}{Bowen Wang}, \bibinfo{person}{Chenhui Zhang}, \bibinfo{person}{Da Yin}, \bibinfo{person}{Dan Zhang}, \bibinfo{person}{Diego Rojas}, \bibinfo{person}{Guanyu Feng}, \bibinfo{person}{Hanlin Zhao}, \bibinfo{person}{Hanyu Lai}, \bibinfo{person}{Hao Yu}, \bibinfo{person}{Hongning Wang}, \bibinfo{person}{Jiadai Sun}, \bibinfo{person}{Jiajie Zhang}, \bibinfo{person}{Jiale Cheng}, \bibinfo{person}{Jiayi Gui}, \bibinfo{person}{Jie Tang}, \bibinfo{person}{Jing Zhang}, \bibinfo{person}{Jingyu Sun}, \bibinfo{person}{Juanzi Li}, \bibinfo{person}{Lei Zhao}, \bibinfo{person}{Lindong Wu}, \bibinfo{person}{Lucen Zhong}, \bibinfo{person}{Mingdao Liu}, \bibinfo{person}{Minlie Huang}, \bibinfo{person}{Peng Zhang}, \bibinfo{person}{Qinkai Zheng}, \bibinfo{person}{Rui Lu}, \bibinfo{person}{Shuaiqi Duan}, \bibinfo{person}{Shudan Zhang}, \bibinfo{person}{Shulin Cao}, \bibinfo{person}{Shuxun
  Yang}, \bibinfo{person}{Weng~Lam Tam}, \bibinfo{person}{Wenyi Zhao}, \bibinfo{person}{Xiao Liu}, \bibinfo{person}{Xiao Xia}, \bibinfo{person}{Xiaohan Zhang}, \bibinfo{person}{Xiaotao Gu}, \bibinfo{person}{Xin Lv}, \bibinfo{person}{Xinghan Liu}, \bibinfo{person}{Xinyi Liu}, \bibinfo{person}{Xinyue Yang}, \bibinfo{person}{Xixuan Song}, \bibinfo{person}{Xunkai Zhang}, \bibinfo{person}{Yifan An}, \bibinfo{person}{Yifan Xu}, \bibinfo{person}{Yilin Niu}, \bibinfo{person}{Yuantao Yang}, \bibinfo{person}{Yueyan Li}, \bibinfo{person}{Yushi Bai}, \bibinfo{person}{Yuxiao Dong}, \bibinfo{person}{Zehan Qi}, \bibinfo{person}{Zhaoyu Wang}, \bibinfo{person}{Zhen Yang}, \bibinfo{person}{Zhengxiao Du}, \bibinfo{person}{Zhenyu Hou}, {and} \bibinfo{person}{Zihan Wang}.} \bibinfo{year}{2024}\natexlab{}.
\newblock \bibinfo{title}{ChatGLM: A Family of Large Language Models from GLM-130B to GLM-4 All Tools}.
\newblock
\showeprint[arxiv]{2406.12793}~[cs.CL]
\urldef\tempurl%
\url{https://arxiv.org/abs/2406.12793}
\showURL{%
\tempurl}


\bibitem[Guo et~al\mbox{.}(2024)]%
        {LR2PPO}
\bibfield{author}{\bibinfo{person}{Taian Guo}, \bibinfo{person}{Taolin Zhang}, \bibinfo{person}{Haoqian Wu}, \bibinfo{person}{Hanjun Li}, \bibinfo{person}{Ruizhi Qiao}, {and} \bibinfo{person}{Xing Sun}.} \bibinfo{year}{2024}\natexlab{}.
\newblock \bibinfo{title}{Multimodal Label Relevance Ranking via Reinforcement Learning}.
\newblock
\showeprint[arxiv]{2407.13221}~[cs.CV]
\urldef\tempurl%
\url{https://arxiv.org/abs/2407.13221}
\showURL{%
\tempurl}


\bibitem[Huang et~al\mbox{.}(2025)]%
        {Vision-R1}
\bibfield{author}{\bibinfo{person}{Wenxuan Huang}, \bibinfo{person}{Bohan Jia}, \bibinfo{person}{Zijie Zhai}, \bibinfo{person}{Shaosheng Cao}, \bibinfo{person}{Zheyu Ye}, \bibinfo{person}{Fei Zhao}, \bibinfo{person}{Zhe Xu}, \bibinfo{person}{Yao Hu}, {and} \bibinfo{person}{Shaohui Lin}.} \bibinfo{year}{2025}\natexlab{}.
\newblock \bibinfo{title}{Vision-R1: Incentivizing Reasoning Capability in Multimodal Large Language Models}.
\newblock
\showeprint[arxiv]{2503.06749}~[cs.CV]
\urldef\tempurl%
\url{https://arxiv.org/abs/2503.06749}
\showURL{%
\tempurl}


\bibitem[Jia et~al\mbox{.}(2021)]%
        {ALIGN}
\bibfield{author}{\bibinfo{person}{Chao Jia}, \bibinfo{person}{Yinfei Yang}, \bibinfo{person}{Ye Xia}, \bibinfo{person}{Yi-Ting Chen}, \bibinfo{person}{Zarana Parekh}, \bibinfo{person}{Hieu Pham}, \bibinfo{person}{Quoc~V. Le}, \bibinfo{person}{Yunhsuan Sung}, \bibinfo{person}{Zhen Li}, {and} \bibinfo{person}{Tom Duerig}.} \bibinfo{year}{2021}\natexlab{}.
\newblock \bibinfo{title}{Scaling Up Visual and Vision-Language Representation Learning With Noisy Text Supervision}.
\newblock
\showeprint[arxiv]{2102.05918}~[cs.CV]
\urldef\tempurl%
\url{https://arxiv.org/abs/2102.05918}
\showURL{%
\tempurl}


\bibitem[Li et~al\mbox{.}(2023)]%
        {BLIP-2}
\bibfield{author}{\bibinfo{person}{Junnan Li}, \bibinfo{person}{Dongxu Li}, \bibinfo{person}{Silvio Savarese}, {and} \bibinfo{person}{Steven Hoi}.} \bibinfo{year}{2023}\natexlab{}.
\newblock \bibinfo{title}{BLIP-2: Bootstrapping Language-Image Pre-training with Frozen Image Encoders and Large Language Models}.
\newblock
\showeprint[arxiv]{2301.12597}~[cs.CV]
\urldef\tempurl%
\url{https://arxiv.org/abs/2301.12597}
\showURL{%
\tempurl}


\bibitem[Li et~al\mbox{.}(2022a)]%
        {BLIP}
\bibfield{author}{\bibinfo{person}{Junnan Li}, \bibinfo{person}{Dongxu Li}, \bibinfo{person}{Caiming Xiong}, {and} \bibinfo{person}{Steven Hoi}.} \bibinfo{year}{2022}\natexlab{a}.
\newblock \bibinfo{title}{BLIP: Bootstrapping Language-Image Pre-training for Unified Vision-Language Understanding and Generation}.
\newblock
\showeprint[arxiv]{2201.12086}~[cs.CV]
\urldef\tempurl%
\url{https://arxiv.org/abs/2201.12086}
\showURL{%
\tempurl}


\bibitem[Li et~al\mbox{.}(2021)]%
        {ALBEF}
\bibfield{author}{\bibinfo{person}{Junnan Li}, \bibinfo{person}{Ramprasaath~R. Selvaraju}, \bibinfo{person}{Akhilesh~Deepak Gotmare}, \bibinfo{person}{Shafiq Joty}, \bibinfo{person}{Caiming Xiong}, {and} \bibinfo{person}{Steven Hoi}.} \bibinfo{year}{2021}\natexlab{}.
\newblock \bibinfo{title}{Align before Fuse: Vision and Language Representation Learning with Momentum Distillation}.
\newblock
\showeprint[arxiv]{2107.07651}~[cs.CV]
\urldef\tempurl%
\url{https://arxiv.org/abs/2107.07651}
\showURL{%
\tempurl}


\bibitem[Li et~al\mbox{.}(2024)]%
        {VideoChat}
\bibfield{author}{\bibinfo{person}{KunChang Li}, \bibinfo{person}{Yinan He}, \bibinfo{person}{Yi Wang}, \bibinfo{person}{Yizhuo Li}, \bibinfo{person}{Wenhai Wang}, \bibinfo{person}{Ping Luo}, \bibinfo{person}{Yali Wang}, \bibinfo{person}{Limin Wang}, {and} \bibinfo{person}{Yu Qiao}.} \bibinfo{year}{2024}\natexlab{}.
\newblock \bibinfo{title}{VideoChat: Chat-Centric Video Understanding}.
\newblock
\showeprint[arxiv]{2305.06355}~[cs.CV]
\urldef\tempurl%
\url{https://arxiv.org/abs/2305.06355}
\showURL{%
\tempurl}


\bibitem[Li et~al\mbox{.}(2022c)]%
        {GLIP}
\bibfield{author}{\bibinfo{person}{Liunian~Harold Li}, \bibinfo{person}{Pengchuan Zhang}, \bibinfo{person}{Haotian Zhang}, \bibinfo{person}{Jianwei Yang}, \bibinfo{person}{Chunyuan Li}, \bibinfo{person}{Yiwu Zhong}, \bibinfo{person}{Lijuan Wang}, \bibinfo{person}{Lu Yuan}, \bibinfo{person}{Lei Zhang}, \bibinfo{person}{Jenq-Neng Hwang}, \bibinfo{person}{Kai-Wei Chang}, {and} \bibinfo{person}{Jianfeng Gao}.} \bibinfo{year}{2022}\natexlab{c}.
\newblock \bibinfo{title}{Grounded Language-Image Pre-training}.
\newblock
\showeprint[arxiv]{2112.03857}~[cs.CV]
\urldef\tempurl%
\url{https://arxiv.org/abs/2112.03857}
\showURL{%
\tempurl}


\bibitem[Li et~al\mbox{.}(2022b)]%
        {DeCLIP}
\bibfield{author}{\bibinfo{person}{Yangguang Li}, \bibinfo{person}{Feng Liang}, \bibinfo{person}{Lichen Zhao}, \bibinfo{person}{Yufeng Cui}, \bibinfo{person}{Wanli Ouyang}, \bibinfo{person}{Jing Shao}, \bibinfo{person}{Fengwei Yu}, {and} \bibinfo{person}{Junjie Yan}.} \bibinfo{year}{2022}\natexlab{b}.
\newblock \bibinfo{title}{Supervision Exists Everywhere: A Data Efficient Contrastive Language-Image Pre-training Paradigm}.
\newblock
\showeprint[arxiv]{2110.05208}~[cs.CV]
\urldef\tempurl%
\url{https://arxiv.org/abs/2110.05208}
\showURL{%
\tempurl}


\bibitem[Liu et~al\mbox{.}(2024b)]%
        {LLaVA-NeXT}
\bibfield{author}{\bibinfo{person}{Haotian Liu}, \bibinfo{person}{Chunyuan Li}, \bibinfo{person}{Yuheng Li}, \bibinfo{person}{Bo Li}, \bibinfo{person}{Yuanhan Zhang}, \bibinfo{person}{Sheng Shen}, {and} \bibinfo{person}{Yong~Jae Lee}.} \bibinfo{year}{2024}\natexlab{b}.
\newblock \bibinfo{title}{LLaVA-NeXT: Improved reasoning, OCR, and world knowledge}.
\newblock
\urldef\tempurl%
\url{https://llava-vl.github.io/blog/2024-01-30-llava-next/}
\showURL{%
\tempurl}


\bibitem[Liu et~al\mbox{.}(2023b)]%
        {LLaVA}
\bibfield{author}{\bibinfo{person}{Haotian Liu}, \bibinfo{person}{Chunyuan Li}, \bibinfo{person}{Qingyang Wu}, {and} \bibinfo{person}{Yong~Jae Lee}.} \bibinfo{year}{2023}\natexlab{b}.
\newblock \bibinfo{title}{Visual Instruction Tuning}.
\newblock
\showeprint[arxiv]{2304.08485}~[cs.CV]
\urldef\tempurl%
\url{https://arxiv.org/abs/2304.08485}
\showURL{%
\tempurl}


\bibitem[Liu et~al\mbox{.}(2023a)]%
        {STAN}
\bibfield{author}{\bibinfo{person}{Ruyang Liu}, \bibinfo{person}{Jingjia Huang}, \bibinfo{person}{Ge Li}, \bibinfo{person}{Jiashi Feng}, \bibinfo{person}{Xinglong Wu}, {and} \bibinfo{person}{Thomas~H. Li}.} \bibinfo{year}{2023}\natexlab{a}.
\newblock \bibinfo{title}{Revisiting Temporal Modeling for CLIP-based Image-to-Video Knowledge Transferring}.
\newblock
\showeprint[arxiv]{2301.11116}~[cs.CV]
\urldef\tempurl%
\url{https://arxiv.org/abs/2301.11116}
\showURL{%
\tempurl}


\bibitem[Liu et~al\mbox{.}(2024a)]%
        {MMBench}
\bibfield{author}{\bibinfo{person}{Yuan Liu}, \bibinfo{person}{Haodong Duan}, \bibinfo{person}{Yuanhan Zhang}, \bibinfo{person}{Bo Li}, \bibinfo{person}{Songyang Zhang}, \bibinfo{person}{Wangbo Zhao}, \bibinfo{person}{Yike Yuan}, \bibinfo{person}{Jiaqi Wang}, \bibinfo{person}{Conghui He}, \bibinfo{person}{Ziwei Liu}, \bibinfo{person}{Kai Chen}, {and} \bibinfo{person}{Dahua Lin}.} \bibinfo{year}{2024}\natexlab{a}.
\newblock \bibinfo{title}{MMBench: Is Your Multi-modal Model an All-around Player?}
\newblock
\showeprint[arxiv]{2307.06281}~[cs.CV]
\urldef\tempurl%
\url{https://arxiv.org/abs/2307.06281}
\showURL{%
\tempurl}


\bibitem[Liu et~al\mbox{.}(2021)]%
        {Que2Search}
\bibfield{author}{\bibinfo{person}{Yiqun Liu}, \bibinfo{person}{Kaushik Rangadurai}, \bibinfo{person}{Yunzhong He}, \bibinfo{person}{Siddarth Malreddy}, \bibinfo{person}{Xunlong Gui}, \bibinfo{person}{Xiaoyi Liu}, {and} \bibinfo{person}{Fedor Borisyuk}.} \bibinfo{year}{2021}\natexlab{}.
\newblock \showarticletitle{Que2Search: Fast and Accurate Query and Document Understanding for Search at Facebook}. In \bibinfo{booktitle}{\emph{Proceedings of the 27th ACM SIGKDD Conference on Knowledge Discovery \& Data Mining}} (Virtual Event, Singapore) \emph{(\bibinfo{series}{KDD '21})}. \bibinfo{publisher}{Association for Computing Machinery}, \bibinfo{address}{New York, NY, USA}, \bibinfo{pages}{3376–3384}.
\newblock
\showISBNx{9781450383325}
\href{https://doi.org/10.1145/3447548.3467127}{doi:\nolinkurl{10.1145/3447548.3467127}}


\bibitem[Liu et~al\mbox{.}(2025)]%
        {Visual-RFT}
\bibfield{author}{\bibinfo{person}{Ziyu Liu}, \bibinfo{person}{Zeyi Sun}, \bibinfo{person}{Yuhang Zang}, \bibinfo{person}{Xiaoyi Dong}, \bibinfo{person}{Yuhang Cao}, \bibinfo{person}{Haodong Duan}, \bibinfo{person}{Dahua Lin}, {and} \bibinfo{person}{Jiaqi Wang}.} \bibinfo{year}{2025}\natexlab{}.
\newblock \bibinfo{title}{Visual-RFT: Visual Reinforcement Fine-Tuning}.
\newblock
\showeprint[arxiv]{2503.01785}~[cs.CV]
\urldef\tempurl%
\url{https://arxiv.org/abs/2503.01785}
\showURL{%
\tempurl}


\bibitem[Lu et~al\mbox{.}(2024)]%
        {DeepSeek-VL}
\bibfield{author}{\bibinfo{person}{Haoyu Lu}, \bibinfo{person}{Wen Liu}, \bibinfo{person}{Bo Zhang}, \bibinfo{person}{Bingxuan Wang}, \bibinfo{person}{Kai Dong}, \bibinfo{person}{Bo Liu}, \bibinfo{person}{Jingxiang Sun}, \bibinfo{person}{Tongzheng Ren}, \bibinfo{person}{Zhuoshu Li}, \bibinfo{person}{Hao Yang}, \bibinfo{person}{Yaofeng Sun}, \bibinfo{person}{Chengqi Deng}, \bibinfo{person}{Hanwei Xu}, \bibinfo{person}{Zhenda Xie}, {and} \bibinfo{person}{Chong Ruan}.} \bibinfo{year}{2024}\natexlab{}.
\newblock \bibinfo{title}{DeepSeek-VL: Towards Real-World Vision-Language Understanding}.
\newblock
\showeprint[arxiv]{2403.05525}~[cs.AI]
\urldef\tempurl%
\url{https://arxiv.org/abs/2403.05525}
\showURL{%
\tempurl}


\bibitem[Maaz et~al\mbox{.}(2024)]%
        {Video-ChatGPT}
\bibfield{author}{\bibinfo{person}{Muhammad Maaz}, \bibinfo{person}{Hanoona Rasheed}, \bibinfo{person}{Salman Khan}, {and} \bibinfo{person}{Fahad~Shahbaz Khan}.} \bibinfo{year}{2024}\natexlab{}.
\newblock \bibinfo{title}{Video-ChatGPT: Towards Detailed Video Understanding via Large Vision and Language Models}.
\newblock
\showeprint[arxiv]{2306.05424}~[cs.CV]
\urldef\tempurl%
\url{https://arxiv.org/abs/2306.05424}
\showURL{%
\tempurl}


\bibitem[Masry et~al\mbox{.}(2022)]%
        {ChartQA}
\bibfield{author}{\bibinfo{person}{Ahmed Masry}, \bibinfo{person}{Do~Xuan Long}, \bibinfo{person}{Jia~Qing Tan}, \bibinfo{person}{Shafiq Joty}, {and} \bibinfo{person}{Enamul Hoque}.} \bibinfo{year}{2022}\natexlab{}.
\newblock \bibinfo{title}{ChartQA: A Benchmark for Question Answering about Charts with Visual and Logical Reasoning}.
\newblock
\showeprint[arxiv]{2203.10244}~[cs.CL]
\urldef\tempurl%
\url{https://arxiv.org/abs/2203.10244}
\showURL{%
\tempurl}


\bibitem[Mathew et~al\mbox{.}(2021)]%
        {DocVQA}
\bibfield{author}{\bibinfo{person}{Minesh Mathew}, \bibinfo{person}{Dimosthenis Karatzas}, {and} \bibinfo{person}{C.~V. Jawahar}.} \bibinfo{year}{2021}\natexlab{}.
\newblock \bibinfo{title}{DocVQA: A Dataset for VQA on Document Images}.
\newblock
\showeprint[arxiv]{2007.00398}~[cs.CV]
\urldef\tempurl%
\url{https://arxiv.org/abs/2007.00398}
\showURL{%
\tempurl}


\bibitem[Nie et~al\mbox{.}({[n.\,d.]})]%
        {videoqformer}
\bibfield{author}{\bibinfo{person}{Yuxiang Nie}, \bibinfo{person}{Han Wang}, \bibinfo{person}{Yanjie Wang}, \bibinfo{person}{Can Huang}, \bibinfo{person}{Liang Lin}, {and} \bibinfo{person}{Guanbin Li}.} \bibinfo{year}{[n.\,d.]}\natexlab{}.
\newblock \showarticletitle{Video Q-Former: Multimodal Large Language Model with Spatio-Temporal Querying Transformer Towards Video Understanding}.
\newblock  (\bibinfo{year}{[n.\,d.]}).
\newblock


\bibitem[OpenAI(2024a)]%
        {GPT-4}
\bibfield{author}{\bibinfo{person}{OpenAI}.} \bibinfo{year}{2024}\natexlab{a}.
\newblock \bibinfo{title}{GPT-4 Technical Report}.
\newblock
\showeprint[arxiv]{2303.08774}~[cs.CL]
\urldef\tempurl%
\url{https://arxiv.org/abs/2303.08774}
\showURL{%
\tempurl}


\bibitem[OpenAI(2024b)]%
        {OpenAI_o1}
\bibfield{author}{\bibinfo{person}{OpenAI}.} \bibinfo{year}{2024}\natexlab{b}.
\newblock \bibinfo{title}{Introducing OpenAI o1}.
\newblock
\urldef\tempurl%
\url{https://openai.com/o1/}
\showURL{%
\tempurl}


\bibitem[Ouyang et~al\mbox{.}(2022)]%
        {InstructGPT}
\bibfield{author}{\bibinfo{person}{Long Ouyang}, \bibinfo{person}{Jeff Wu}, \bibinfo{person}{Xu Jiang}, \bibinfo{person}{Diogo Almeida}, \bibinfo{person}{Carroll~L. Wainwright}, \bibinfo{person}{Pamela Mishkin}, \bibinfo{person}{Chong Zhang}, \bibinfo{person}{Sandhini Agarwal}, \bibinfo{person}{Katarina Slama}, \bibinfo{person}{Alex Ray}, \bibinfo{person}{John Schulman}, \bibinfo{person}{Jacob Hilton}, \bibinfo{person}{Fraser Kelton}, \bibinfo{person}{Luke Miller}, \bibinfo{person}{Maddie Simens}, \bibinfo{person}{Amanda Askell}, \bibinfo{person}{Peter Welinder}, \bibinfo{person}{Paul Christiano}, \bibinfo{person}{Jan Leike}, {and} \bibinfo{person}{Ryan Lowe}.} \bibinfo{year}{2022}\natexlab{}.
\newblock \bibinfo{title}{Training language models to follow instructions with human feedback}.
\newblock
\showeprint[arxiv]{2203.02155}~[cs.CL]
\urldef\tempurl%
\url{https://arxiv.org/abs/2203.02155}
\showURL{%
\tempurl}


\bibitem[Paiss et~al\mbox{.}(2023)]%
        {CountBench}
\bibfield{author}{\bibinfo{person}{Roni Paiss}, \bibinfo{person}{Ariel Ephrat}, \bibinfo{person}{Omer Tov}, \bibinfo{person}{Shiran Zada}, \bibinfo{person}{Inbar Mosseri}, \bibinfo{person}{Michal Irani}, {and} \bibinfo{person}{Tali Dekel}.} \bibinfo{year}{2023}\natexlab{}.
\newblock \bibinfo{title}{Teaching CLIP to Count to Ten}.
\newblock
\showeprint[arxiv]{2302.12066}~[cs.CV]
\urldef\tempurl%
\url{https://arxiv.org/abs/2302.12066}
\showURL{%
\tempurl}


\bibitem[Papineni et~al\mbox{.}(2002)]%
        {BLEU}
\bibfield{author}{\bibinfo{person}{Kishore Papineni}, \bibinfo{person}{Salim Roukos}, \bibinfo{person}{Todd Ward}, {and} \bibinfo{person}{Wei-Jing Zhu}.} \bibinfo{year}{2002}\natexlab{}.
\newblock \showarticletitle{Bleu: a method for automatic evaluation of machine translation}. In \bibinfo{booktitle}{\emph{Proceedings of the 40th annual meeting of the Association for Computational Linguistics}}. \bibinfo{pages}{311--318}.
\newblock


\bibitem[Peng et~al\mbox{.}(2025)]%
        {Skywork-R1V}
\bibfield{author}{\bibinfo{person}{Yi Peng}, \bibinfo{person}{Chris}, \bibinfo{person}{Xiaokun Wang}, \bibinfo{person}{Yichen Wei}, \bibinfo{person}{Jiangbo Pei}, \bibinfo{person}{Weijie Qiu}, \bibinfo{person}{Ai Jian}, \bibinfo{person}{Yunzhuo Hao}, \bibinfo{person}{Jiachun Pan}, \bibinfo{person}{Tianyidan Xie}, \bibinfo{person}{Li Ge}, \bibinfo{person}{Rongxian Zhuang}, \bibinfo{person}{Xuchen Song}, \bibinfo{person}{Yang Liu}, {and} \bibinfo{person}{Yahui Zhou}.} \bibinfo{year}{2025}\natexlab{}.
\newblock \bibinfo{title}{Skywork R1V: Pioneering Multimodal Reasoning with Chain-of-Thought}.
\newblock
\showeprint[arxiv]{2504.05599}~[cs.CV]
\urldef\tempurl%
\url{https://arxiv.org/abs/2504.05599}
\showURL{%
\tempurl}


\bibitem[Qwen et~al\mbox{.}(2025)]%
        {Qwen2_5}
\bibfield{author}{\bibinfo{person}{Qwen}, \bibinfo{person}{:}, \bibinfo{person}{An Yang}, \bibinfo{person}{Baosong Yang}, \bibinfo{person}{Beichen Zhang}, \bibinfo{person}{Binyuan Hui}, \bibinfo{person}{Bo Zheng}, \bibinfo{person}{Bowen Yu}, \bibinfo{person}{Chengyuan Li}, \bibinfo{person}{Dayiheng Liu}, \bibinfo{person}{Fei Huang}, \bibinfo{person}{Haoran Wei}, \bibinfo{person}{Huan Lin}, \bibinfo{person}{Jian Yang}, \bibinfo{person}{Jianhong Tu}, \bibinfo{person}{Jianwei Zhang}, \bibinfo{person}{Jianxin Yang}, \bibinfo{person}{Jiaxi Yang}, \bibinfo{person}{Jingren Zhou}, \bibinfo{person}{Junyang Lin}, \bibinfo{person}{Kai Dang}, \bibinfo{person}{Keming Lu}, \bibinfo{person}{Keqin Bao}, \bibinfo{person}{Kexin Yang}, \bibinfo{person}{Le Yu}, \bibinfo{person}{Mei Li}, \bibinfo{person}{Mingfeng Xue}, \bibinfo{person}{Pei Zhang}, \bibinfo{person}{Qin Zhu}, \bibinfo{person}{Rui Men}, \bibinfo{person}{Runji Lin}, \bibinfo{person}{Tianhao Li}, \bibinfo{person}{Tianyi Tang}, \bibinfo{person}{Tingyu Xia},
  \bibinfo{person}{Xingzhang Ren}, \bibinfo{person}{Xuancheng Ren}, \bibinfo{person}{Yang Fan}, \bibinfo{person}{Yang Su}, \bibinfo{person}{Yichang Zhang}, \bibinfo{person}{Yu Wan}, \bibinfo{person}{Yuqiong Liu}, \bibinfo{person}{Zeyu Cui}, \bibinfo{person}{Zhenru Zhang}, {and} \bibinfo{person}{Zihan Qiu}.} \bibinfo{year}{2025}\natexlab{}.
\newblock \bibinfo{title}{Qwen2.5 Technical Report}.
\newblock
\showeprint[arxiv]{2412.15115}~[cs.CL]
\urldef\tempurl%
\url{https://arxiv.org/abs/2412.15115}
\showURL{%
\tempurl}


\bibitem[Radford et~al\mbox{.}(2021)]%
        {CLIP}
\bibfield{author}{\bibinfo{person}{Alec Radford}, \bibinfo{person}{Jong~Wook Kim}, \bibinfo{person}{Chris Hallacy}, \bibinfo{person}{Aditya Ramesh}, \bibinfo{person}{Gabriel Goh}, \bibinfo{person}{Sandhini Agarwal}, \bibinfo{person}{Girish Sastry}, \bibinfo{person}{Amanda Askell}, \bibinfo{person}{Pamela Mishkin}, \bibinfo{person}{Jack Clark}, \bibinfo{person}{Gretchen Krueger}, {and} \bibinfo{person}{Ilya Sutskever}.} \bibinfo{year}{2021}\natexlab{}.
\newblock \bibinfo{title}{Learning Transferable Visual Models From Natural Language Supervision}.
\newblock
\showeprint[arxiv]{2103.00020}~[cs.CV]
\urldef\tempurl%
\url{https://arxiv.org/abs/2103.00020}
\showURL{%
\tempurl}


\bibitem[Schulman et~al\mbox{.}(2017)]%
        {PPO}
\bibfield{author}{\bibinfo{person}{John Schulman}, \bibinfo{person}{Filip Wolski}, \bibinfo{person}{Prafulla Dhariwal}, \bibinfo{person}{Alec Radford}, {and} \bibinfo{person}{Oleg Klimov}.} \bibinfo{year}{2017}\natexlab{}.
\newblock \bibinfo{title}{Proximal Policy Optimization Algorithms}.
\newblock
\showeprint[arxiv]{1707.06347}~[cs.LG]
\urldef\tempurl%
\url{https://arxiv.org/abs/1707.06347}
\showURL{%
\tempurl}


\bibitem[Shao et~al\mbox{.}(2024)]%
        {DeepSeek-Math}
\bibfield{author}{\bibinfo{person}{Zhihong Shao}, \bibinfo{person}{Peiyi Wang}, \bibinfo{person}{Qihao Zhu}, \bibinfo{person}{Runxin Xu}, \bibinfo{person}{Junxiao Song}, \bibinfo{person}{Xiao Bi}, \bibinfo{person}{Haowei Zhang}, \bibinfo{person}{Mingchuan Zhang}, \bibinfo{person}{Y.~K. Li}, \bibinfo{person}{Y. Wu}, {and} \bibinfo{person}{Daya Guo}.} \bibinfo{year}{2024}\natexlab{}.
\newblock \bibinfo{title}{DeepSeekMath: Pushing the Limits of Mathematical Reasoning in Open Language Models}.
\newblock
\showeprint[arxiv]{2402.03300}~[cs.CL]
\urldef\tempurl%
\url{https://arxiv.org/abs/2402.03300}
\showURL{%
\tempurl}


\bibitem[Sheng et~al\mbox{.}(2024)]%
        {verl}
\bibfield{author}{\bibinfo{person}{Guangming Sheng}, \bibinfo{person}{Chi Zhang}, \bibinfo{person}{Zilingfeng Ye}, \bibinfo{person}{Xibin Wu}, \bibinfo{person}{Wang Zhang}, \bibinfo{person}{Ru Zhang}, \bibinfo{person}{Yanghua Peng}, \bibinfo{person}{Haibin Lin}, {and} \bibinfo{person}{Chuan Wu}.} \bibinfo{year}{2024}\natexlab{}.
\newblock \showarticletitle{HybridFlow: A Flexible and Efficient RLHF Framework}.
\newblock \bibinfo{journal}{\emph{arXiv preprint arXiv: 2409.19256}} (\bibinfo{year}{2024}).
\newblock


\bibitem[Team et~al\mbox{.}(2025)]%
        {Kimi-VL}
\bibfield{author}{\bibinfo{person}{Kimi Team}, \bibinfo{person}{Angang Du}, \bibinfo{person}{Bohong Yin}, \bibinfo{person}{Bowei Xing}, \bibinfo{person}{Bowen Qu}, \bibinfo{person}{Bowen Wang}, \bibinfo{person}{Cheng Chen}, \bibinfo{person}{Chenlin Zhang}, \bibinfo{person}{Chenzhuang Du}, \bibinfo{person}{Chu Wei}, \bibinfo{person}{Congcong Wang}, \bibinfo{person}{Dehao Zhang}, \bibinfo{person}{Dikang Du}, \bibinfo{person}{Dongliang Wang}, \bibinfo{person}{Enming Yuan}, \bibinfo{person}{Enzhe Lu}, \bibinfo{person}{Fang Li}, \bibinfo{person}{Flood Sung}, \bibinfo{person}{Guangda Wei}, \bibinfo{person}{Guokun Lai}, \bibinfo{person}{Han Zhu}, \bibinfo{person}{Hao Ding}, \bibinfo{person}{Hao Hu}, \bibinfo{person}{Hao Yang}, \bibinfo{person}{Hao Zhang}, \bibinfo{person}{Haoning Wu}, \bibinfo{person}{Haotian Yao}, \bibinfo{person}{Haoyu Lu}, \bibinfo{person}{Heng Wang}, \bibinfo{person}{Hongcheng Gao}, \bibinfo{person}{Huabin Zheng}, \bibinfo{person}{Jiaming Li}, \bibinfo{person}{Jianlin Su},
  \bibinfo{person}{Jianzhou Wang}, \bibinfo{person}{Jiaqi Deng}, \bibinfo{person}{Jiezhong Qiu}, \bibinfo{person}{Jin Xie}, \bibinfo{person}{Jinhong Wang}, \bibinfo{person}{Jingyuan Liu}, \bibinfo{person}{Junjie Yan}, \bibinfo{person}{Kun Ouyang}, \bibinfo{person}{Liang Chen}, \bibinfo{person}{Lin Sui}, \bibinfo{person}{Longhui Yu}, \bibinfo{person}{Mengfan Dong}, \bibinfo{person}{Mengnan Dong}, \bibinfo{person}{Nuo Xu}, \bibinfo{person}{Pengyu Cheng}, \bibinfo{person}{Qizheng Gu}, \bibinfo{person}{Runjie Zhou}, \bibinfo{person}{Shaowei Liu}, \bibinfo{person}{Sihan Cao}, \bibinfo{person}{Tao Yu}, \bibinfo{person}{Tianhui Song}, \bibinfo{person}{Tongtong Bai}, \bibinfo{person}{Wei Song}, \bibinfo{person}{Weiran He}, \bibinfo{person}{Weixiao Huang}, \bibinfo{person}{Weixin Xu}, \bibinfo{person}{Xiaokun Yuan}, \bibinfo{person}{Xingcheng Yao}, \bibinfo{person}{Xingzhe Wu}, \bibinfo{person}{Xinxing Zu}, \bibinfo{person}{Xinyu Zhou}, \bibinfo{person}{Xinyuan Wang}, \bibinfo{person}{Y. Charles}, \bibinfo{person}{Yan
  Zhong}, \bibinfo{person}{Yang Li}, \bibinfo{person}{Yangyang Hu}, \bibinfo{person}{Yanru Chen}, \bibinfo{person}{Yejie Wang}, \bibinfo{person}{Yibo Liu}, \bibinfo{person}{Yibo Miao}, \bibinfo{person}{Yidao Qin}, \bibinfo{person}{Yimin Chen}, \bibinfo{person}{Yiping Bao}, \bibinfo{person}{Yiqin Wang}, \bibinfo{person}{Yongsheng Kang}, \bibinfo{person}{Yuanxin Liu}, \bibinfo{person}{Yulun Du}, \bibinfo{person}{Yuxin Wu}, \bibinfo{person}{Yuzhi Wang}, \bibinfo{person}{Yuzi Yan}, \bibinfo{person}{Zaida Zhou}, \bibinfo{person}{Zhaowei Li}, \bibinfo{person}{Zhejun Jiang}, \bibinfo{person}{Zheng Zhang}, \bibinfo{person}{Zhilin Yang}, \bibinfo{person}{Zhiqi Huang}, \bibinfo{person}{Zihao Huang}, \bibinfo{person}{Zijia Zhao}, \bibinfo{person}{Ziwei Chen}, {and} \bibinfo{person}{Zongyu Lin}.} \bibinfo{year}{2025}\natexlab{}.
\newblock \bibinfo{title}{Kimi-VL Technical Report}.
\newblock
\showeprint[arxiv]{2504.07491}~[cs.CV]
\urldef\tempurl%
\url{https://arxiv.org/abs/2504.07491}
\showURL{%
\tempurl}


\bibitem[Touvron et~al\mbox{.}(2023)]%
        {Llama2}
\bibfield{author}{\bibinfo{person}{Hugo Touvron}, \bibinfo{person}{Louis Martin}, \bibinfo{person}{Kevin Stone}, \bibinfo{person}{Peter Albert}, \bibinfo{person}{Amjad Almahairi}, \bibinfo{person}{Yasmine Babaei}, \bibinfo{person}{Nikolay Bashlykov}, \bibinfo{person}{Soumya Batra}, \bibinfo{person}{Prajjwal Bhargava}, \bibinfo{person}{Shruti Bhosale}, \bibinfo{person}{Dan Bikel}, \bibinfo{person}{Lukas Blecher}, \bibinfo{person}{Cristian~Canton Ferrer}, \bibinfo{person}{Moya Chen}, \bibinfo{person}{Guillem Cucurull}, \bibinfo{person}{David Esiobu}, \bibinfo{person}{Jude Fernandes}, \bibinfo{person}{Jeremy Fu}, \bibinfo{person}{Wenyin Fu}, \bibinfo{person}{Brian Fuller}, \bibinfo{person}{Cynthia Gao}, \bibinfo{person}{Vedanuj Goswami}, \bibinfo{person}{Naman Goyal}, \bibinfo{person}{Anthony Hartshorn}, \bibinfo{person}{Saghar Hosseini}, \bibinfo{person}{Rui Hou}, \bibinfo{person}{Hakan Inan}, \bibinfo{person}{Marcin Kardas}, \bibinfo{person}{Viktor Kerkez}, \bibinfo{person}{Madian Khabsa},
  \bibinfo{person}{Isabel Kloumann}, \bibinfo{person}{Artem Korenev}, \bibinfo{person}{Punit~Singh Koura}, \bibinfo{person}{Marie-Anne Lachaux}, \bibinfo{person}{Thibaut Lavril}, \bibinfo{person}{Jenya Lee}, \bibinfo{person}{Diana Liskovich}, \bibinfo{person}{Yinghai Lu}, \bibinfo{person}{Yuning Mao}, \bibinfo{person}{Xavier Martinet}, \bibinfo{person}{Todor Mihaylov}, \bibinfo{person}{Pushkar Mishra}, \bibinfo{person}{Igor Molybog}, \bibinfo{person}{Yixin Nie}, \bibinfo{person}{Andrew Poulton}, \bibinfo{person}{Jeremy Reizenstein}, \bibinfo{person}{Rashi Rungta}, \bibinfo{person}{Kalyan Saladi}, \bibinfo{person}{Alan Schelten}, \bibinfo{person}{Ruan Silva}, \bibinfo{person}{Eric~Michael Smith}, \bibinfo{person}{Ranjan Subramanian}, \bibinfo{person}{Xiaoqing~Ellen Tan}, \bibinfo{person}{Binh Tang}, \bibinfo{person}{Ross Taylor}, \bibinfo{person}{Adina Williams}, \bibinfo{person}{Jian~Xiang Kuan}, \bibinfo{person}{Puxin Xu}, \bibinfo{person}{Zheng Yan}, \bibinfo{person}{Iliyan Zarov}, \bibinfo{person}{Yuchen
  Zhang}, \bibinfo{person}{Angela Fan}, \bibinfo{person}{Melanie Kambadur}, \bibinfo{person}{Sharan Narang}, \bibinfo{person}{Aurelien Rodriguez}, \bibinfo{person}{Robert Stojnic}, \bibinfo{person}{Sergey Edunov}, {and} \bibinfo{person}{Thomas Scialom}.} \bibinfo{year}{2023}\natexlab{}.
\newblock \bibinfo{title}{Llama 2: Open Foundation and Fine-Tuned Chat Models}.
\newblock
\showeprint[arxiv]{2307.09288}~[cs.CL]
\urldef\tempurl%
\url{https://arxiv.org/abs/2307.09288}
\showURL{%
\tempurl}


\bibitem[Wang et~al\mbox{.}(2024c)]%
        {dynamicvlm}
\bibfield{author}{\bibinfo{person}{Han Wang}, \bibinfo{person}{Yuxiang Nie}, \bibinfo{person}{Yongjie Ye}, \bibinfo{person}{Deng GuanYu}, \bibinfo{person}{Yanjie Wang}, \bibinfo{person}{Shuai Li}, \bibinfo{person}{Haiyang Yu}, \bibinfo{person}{Jinghui Lu}, {and} \bibinfo{person}{Can Huang}.} \bibinfo{year}{2024}\natexlab{c}.
\newblock \showarticletitle{Dynamic-vlm: Simple dynamic visual token compression for videollm}.
\newblock \bibinfo{journal}{\emph{arXiv preprint arXiv:2412.09530}} (\bibinfo{year}{2024}).
\newblock


\bibitem[Wang et~al\mbox{.}(2022b)]%
        {ptseformer}
\bibfield{author}{\bibinfo{person}{Han Wang}, \bibinfo{person}{Jun Tang}, \bibinfo{person}{Xiaodong Liu}, \bibinfo{person}{Shanyan Guan}, \bibinfo{person}{Rong Xie}, {and} \bibinfo{person}{Li Song}.} \bibinfo{year}{2022}\natexlab{b}.
\newblock \showarticletitle{Ptseformer: Progressive temporal-spatial enhanced transformer towards video object detection}. In \bibinfo{booktitle}{\emph{European conference on computer vision}}. Springer, \bibinfo{pages}{732--747}.
\newblock


\bibitem[Wang et~al\mbox{.}({[n.\,d.]})]%
        {gloma}
\bibfield{author}{\bibinfo{person}{Han Wang}, \bibinfo{person}{Yanjie Wang}, \bibinfo{person}{Yang Li}, {and} \bibinfo{person}{Can Huang}.} \bibinfo{year}{[n.\,d.]}\natexlab{}.
\newblock \showarticletitle{GLOMA: Global Video Text Spotting with Morphological Association}. In \bibinfo{booktitle}{\emph{The Thirteenth International Conference on Learning Representations}}.
\newblock


\bibitem[Wang et~al\mbox{.}(2025b)]%
        {vora}
\bibfield{author}{\bibinfo{person}{Han Wang}, \bibinfo{person}{Yongjie Ye}, \bibinfo{person}{Bingru Li}, \bibinfo{person}{Yuxiang Nie}, \bibinfo{person}{Jinghui Lu}, \bibinfo{person}{Jingqun Tang}, \bibinfo{person}{Yanjie Wang}, {and} \bibinfo{person}{Can Huang}.} \bibinfo{year}{2025}\natexlab{b}.
\newblock \showarticletitle{Vision as lora}.
\newblock \bibinfo{journal}{\emph{arXiv preprint arXiv:2503.20680}} (\bibinfo{year}{2025}).
\newblock


\bibitem[Wang et~al\mbox{.}(2024d)]%
        {elysium}
\bibfield{author}{\bibinfo{person}{Han Wang}, \bibinfo{person}{Yongjie Ye}, \bibinfo{person}{Yanjie Wang}, \bibinfo{person}{Yuxiang Nie}, {and} \bibinfo{person}{Can Huang}.} \bibinfo{year}{2024}\natexlab{d}.
\newblock \showarticletitle{Elysium: Exploring object-level perception in videos via mllm}. In \bibinfo{booktitle}{\emph{European Conference on Computer Vision}}. Springer, \bibinfo{pages}{166--185}.
\newblock


\bibitem[Wang et~al\mbox{.}(2024a)]%
        {Qwen2-VL}
\bibfield{author}{\bibinfo{person}{Peng Wang}, \bibinfo{person}{Shuai Bai}, \bibinfo{person}{Sinan Tan}, \bibinfo{person}{Shijie Wang}, \bibinfo{person}{Zhihao Fan}, \bibinfo{person}{Jinze Bai}, \bibinfo{person}{Keqin Chen}, \bibinfo{person}{Xuejing Liu}, \bibinfo{person}{Jialin Wang}, \bibinfo{person}{Wenbin Ge}, \bibinfo{person}{Yang Fan}, \bibinfo{person}{Kai Dang}, \bibinfo{person}{Mengfei Du}, \bibinfo{person}{Xuancheng Ren}, \bibinfo{person}{Rui Men}, \bibinfo{person}{Dayiheng Liu}, \bibinfo{person}{Chang Zhou}, \bibinfo{person}{Jingren Zhou}, {and} \bibinfo{person}{Junyang Lin}.} \bibinfo{year}{2024}\natexlab{a}.
\newblock \bibinfo{title}{Qwen2-VL: Enhancing Vision-Language Model's Perception of the World at Any Resolution}.
\newblock
\showeprint[arxiv]{2409.12191}~[cs.CV]
\urldef\tempurl%
\url{https://arxiv.org/abs/2409.12191}
\showURL{%
\tempurl}


\bibitem[Wang et~al\mbox{.}(2024b)]%
        {InternVideo2}
\bibfield{author}{\bibinfo{person}{Yi Wang}, \bibinfo{person}{Kunchang Li}, \bibinfo{person}{Xinhao Li}, \bibinfo{person}{Jiashuo Yu}, \bibinfo{person}{Yinan He}, \bibinfo{person}{Chenting Wang}, \bibinfo{person}{Guo Chen}, \bibinfo{person}{Baoqi Pei}, \bibinfo{person}{Ziang Yan}, \bibinfo{person}{Rongkun Zheng}, \bibinfo{person}{Jilan Xu}, \bibinfo{person}{Zun Wang}, \bibinfo{person}{Yansong Shi}, \bibinfo{person}{Tianxiang Jiang}, \bibinfo{person}{Songze Li}, \bibinfo{person}{Hongjie Zhang}, \bibinfo{person}{Yifei Huang}, \bibinfo{person}{Yu Qiao}, \bibinfo{person}{Yali Wang}, {and} \bibinfo{person}{Limin Wang}.} \bibinfo{year}{2024}\natexlab{b}.
\newblock \bibinfo{title}{InternVideo2: Scaling Foundation Models for Multimodal Video Understanding}.
\newblock
\showeprint[arxiv]{2403.15377}~[cs.CV]
\urldef\tempurl%
\url{https://arxiv.org/abs/2403.15377}
\showURL{%
\tempurl}


\bibitem[Wang et~al\mbox{.}(2022a)]%
        {InternVideo}
\bibfield{author}{\bibinfo{person}{Yi Wang}, \bibinfo{person}{Kunchang Li}, \bibinfo{person}{Yizhuo Li}, \bibinfo{person}{Yinan He}, \bibinfo{person}{Bingkun Huang}, \bibinfo{person}{Zhiyu Zhao}, \bibinfo{person}{Hongjie Zhang}, \bibinfo{person}{Jilan Xu}, \bibinfo{person}{Yi Liu}, \bibinfo{person}{Zun Wang}, \bibinfo{person}{Sen Xing}, \bibinfo{person}{Guo Chen}, \bibinfo{person}{Junting Pan}, \bibinfo{person}{Jiashuo Yu}, \bibinfo{person}{Yali Wang}, \bibinfo{person}{Limin Wang}, {and} \bibinfo{person}{Yu Qiao}.} \bibinfo{year}{2022}\natexlab{a}.
\newblock \bibinfo{title}{InternVideo: General Video Foundation Models via Generative and Discriminative Learning}.
\newblock
\showeprint[arxiv]{2212.03191}~[cs.CV]
\urldef\tempurl%
\url{https://arxiv.org/abs/2212.03191}
\showURL{%
\tempurl}


\bibitem[Wang et~al\mbox{.}(2025a)]%
        {InternVideo2.5}
\bibfield{author}{\bibinfo{person}{Yi Wang}, \bibinfo{person}{Xinhao Li}, \bibinfo{person}{Ziang Yan}, \bibinfo{person}{Yinan He}, \bibinfo{person}{Jiashuo Yu}, \bibinfo{person}{Xiangyu Zeng}, \bibinfo{person}{Chenting Wang}, \bibinfo{person}{Changlian Ma}, \bibinfo{person}{Haian Huang}, \bibinfo{person}{Jianfei Gao}, \bibinfo{person}{Min Dou}, \bibinfo{person}{Kai Chen}, \bibinfo{person}{Wenhai Wang}, \bibinfo{person}{Yu Qiao}, \bibinfo{person}{Yali Wang}, {and} \bibinfo{person}{Limin Wang}.} \bibinfo{year}{2025}\natexlab{a}.
\newblock \bibinfo{title}{InternVideo2.5: Empowering Video MLLMs with Long and Rich Context Modeling}.
\newblock
\showeprint[arxiv]{2501.12386}~[cs.CV]
\urldef\tempurl%
\url{https://arxiv.org/abs/2501.12386}
\showURL{%
\tempurl}


\bibitem[Wen et~al\mbox{.}(2024)]%
        {DQU-CIR}
\bibfield{author}{\bibinfo{person}{Haokun Wen}, \bibinfo{person}{Xuemeng Song}, \bibinfo{person}{Xiaolin Chen}, \bibinfo{person}{Yinwei Wei}, \bibinfo{person}{Liqiang Nie}, {and} \bibinfo{person}{Tat-Seng Chua}.} \bibinfo{year}{2024}\natexlab{}.
\newblock \showarticletitle{Simple but Effective Raw-Data Level Multimodal Fusion for Composed Image Retrieval}. In \bibinfo{booktitle}{\emph{Proceedings of the 47th International ACM SIGIR Conference on Research and Development in Information Retrieval}} \emph{(\bibinfo{series}{SIGIR 2024})}. \bibinfo{publisher}{ACM}, \bibinfo{pages}{229–239}.
\newblock
\href{https://doi.org/10.1145/3626772.3657727}{doi:\nolinkurl{10.1145/3626772.3657727}}


\bibitem[Wen et~al\mbox{.}(2023)]%
        {vican}
\bibfield{author}{\bibinfo{person}{Zhoufutu Wen}, \bibinfo{person}{Xinyu Zhao}, \bibinfo{person}{Zhipeng Jin}, \bibinfo{person}{Yi Yang}, \bibinfo{person}{Wei Jia}, \bibinfo{person}{Xiaodong Chen}, \bibinfo{person}{Shuanglong Li}, {and} \bibinfo{person}{Lin Liu}.} \bibinfo{year}{2023}\natexlab{}.
\newblock \bibinfo{title}{Enhancing Dynamic Image Advertising with Vision-Language Pre-training}.
\newblock
\showeprint[arxiv]{2306.14112}~[cs.IR]
\urldef\tempurl%
\url{https://arxiv.org/abs/2306.14112}
\showURL{%
\tempurl}


\bibitem[Wu et~al\mbox{.}(2024)]%
        {DeepSeek-VL2}
\bibfield{author}{\bibinfo{person}{Zhiyu Wu}, \bibinfo{person}{Xiaokang Chen}, \bibinfo{person}{Zizheng Pan}, \bibinfo{person}{Xingchao Liu}, \bibinfo{person}{Wen Liu}, \bibinfo{person}{Damai Dai}, \bibinfo{person}{Huazuo Gao}, \bibinfo{person}{Yiyang Ma}, \bibinfo{person}{Chengyue Wu}, \bibinfo{person}{Bingxuan Wang}, \bibinfo{person}{Zhenda Xie}, \bibinfo{person}{Yu Wu}, \bibinfo{person}{Kai Hu}, \bibinfo{person}{Jiawei Wang}, \bibinfo{person}{Yaofeng Sun}, \bibinfo{person}{Yukun Li}, \bibinfo{person}{Yishi Piao}, \bibinfo{person}{Kang Guan}, \bibinfo{person}{Aixin Liu}, \bibinfo{person}{Xin Xie}, \bibinfo{person}{Yuxiang You}, \bibinfo{person}{Kai Dong}, \bibinfo{person}{Xingkai Yu}, \bibinfo{person}{Haowei Zhang}, \bibinfo{person}{Liang Zhao}, \bibinfo{person}{Yisong Wang}, {and} \bibinfo{person}{Chong Ruan}.} \bibinfo{year}{2024}\natexlab{}.
\newblock \bibinfo{title}{DeepSeek-VL2: Mixture-of-Experts Vision-Language Models for Advanced Multimodal Understanding}.
\newblock
\showeprint[arxiv]{2412.10302}~[cs.CV]
\urldef\tempurl%
\url{https://arxiv.org/abs/2412.10302}
\showURL{%
\tempurl}


\bibitem[Xu et~al\mbox{.}(2024)]%
        {ARMMT}
\bibfield{author}{\bibinfo{person}{Enqiang Xu}, \bibinfo{person}{Xinhui Li}, \bibinfo{person}{Zhigong Zhou}, \bibinfo{person}{Jiahao Ji}, \bibinfo{person}{Jinyuan Zhao}, \bibinfo{person}{Dadong Miao}, \bibinfo{person}{Songlin Wang}, \bibinfo{person}{Lin Liu}, {and} \bibinfo{person}{Sulong Xu}.} \bibinfo{year}{2024}\natexlab{}.
\newblock \bibinfo{title}{Advancing Re-Ranking with Multimodal Fusion and Target-Oriented Auxiliary Tasks in E-Commerce Search}.
\newblock
\showeprint[arxiv]{2408.05751}~[cs.IR]
\urldef\tempurl%
\url{https://arxiv.org/abs/2408.05751}
\showURL{%
\tempurl}


\bibitem[Yang et~al\mbox{.}(2025)]%
        {Baichuan2}
\bibfield{author}{\bibinfo{person}{Aiyuan Yang}, \bibinfo{person}{Bin Xiao}, \bibinfo{person}{Bingning Wang}, \bibinfo{person}{Borong Zhang}, \bibinfo{person}{Ce Bian}, \bibinfo{person}{Chao Yin}, \bibinfo{person}{Chenxu Lv}, \bibinfo{person}{Da Pan}, \bibinfo{person}{Dian Wang}, \bibinfo{person}{Dong Yan}, \bibinfo{person}{Fan Yang}, \bibinfo{person}{Fei Deng}, \bibinfo{person}{Feng Wang}, \bibinfo{person}{Feng Liu}, \bibinfo{person}{Guangwei Ai}, \bibinfo{person}{Guosheng Dong}, \bibinfo{person}{Haizhou Zhao}, \bibinfo{person}{Hang Xu}, \bibinfo{person}{Haoze Sun}, \bibinfo{person}{Hongda Zhang}, \bibinfo{person}{Hui Liu}, \bibinfo{person}{Jiaming Ji}, \bibinfo{person}{Jian Xie}, \bibinfo{person}{JunTao Dai}, \bibinfo{person}{Kun Fang}, \bibinfo{person}{Lei Su}, \bibinfo{person}{Liang Song}, \bibinfo{person}{Lifeng Liu}, \bibinfo{person}{Liyun Ru}, \bibinfo{person}{Luyao Ma}, \bibinfo{person}{Mang Wang}, \bibinfo{person}{Mickel Liu}, \bibinfo{person}{MingAn Lin}, \bibinfo{person}{Nuolan Nie},
  \bibinfo{person}{Peidong Guo}, \bibinfo{person}{Ruiyang Sun}, \bibinfo{person}{Tao Zhang}, \bibinfo{person}{Tianpeng Li}, \bibinfo{person}{Tianyu Li}, \bibinfo{person}{Wei Cheng}, \bibinfo{person}{Weipeng Chen}, \bibinfo{person}{Xiangrong Zeng}, \bibinfo{person}{Xiaochuan Wang}, \bibinfo{person}{Xiaoxi Chen}, \bibinfo{person}{Xin Men}, \bibinfo{person}{Xin Yu}, \bibinfo{person}{Xuehai Pan}, \bibinfo{person}{Yanjun Shen}, \bibinfo{person}{Yiding Wang}, \bibinfo{person}{Yiyu Li}, \bibinfo{person}{Youxin Jiang}, \bibinfo{person}{Yuchen Gao}, \bibinfo{person}{Yupeng Zhang}, \bibinfo{person}{Zenan Zhou}, {and} \bibinfo{person}{Zhiying Wu}.} \bibinfo{year}{2025}\natexlab{}.
\newblock \bibinfo{title}{Baichuan 2: Open Large-scale Language Models}.
\newblock
\showeprint[arxiv]{2309.10305}~[cs.CL]
\urldef\tempurl%
\url{https://arxiv.org/abs/2309.10305}
\showURL{%
\tempurl}


\bibitem[Yang et~al\mbox{.}(2024)]%
        {LDRE}
\bibfield{author}{\bibinfo{person}{Zhenyu Yang}, \bibinfo{person}{Dizhan Xue}, \bibinfo{person}{Shengsheng Qian}, \bibinfo{person}{Weiming Dong}, {and} \bibinfo{person}{Changsheng Xu}.} \bibinfo{year}{2024}\natexlab{}.
\newblock \showarticletitle{LDRE: LLM-based Divergent Reasoning and Ensemble for Zero-Shot Composed Image Retrieval}. In \bibinfo{booktitle}{\emph{Proceedings of the 47th International ACM SIGIR Conference on Research and Development in Information Retrieval}} (Washington DC, USA) \emph{(\bibinfo{series}{SIGIR '24})}. \bibinfo{publisher}{Association for Computing Machinery}, \bibinfo{address}{New York, NY, USA}, \bibinfo{pages}{80–90}.
\newblock
\showISBNx{9798400704314}
\href{https://doi.org/10.1145/3626772.3657740}{doi:\nolinkurl{10.1145/3626772.3657740}}


\bibitem[Yao et~al\mbox{.}(2021)]%
        {MASM}
\bibfield{author}{\bibinfo{person}{Shaowei Yao}, \bibinfo{person}{Jiwei Tan}, \bibinfo{person}{Xi Chen}, \bibinfo{person}{Keping Yang}, \bibinfo{person}{Rong Xiao}, \bibinfo{person}{Hongbo Deng}, {and} \bibinfo{person}{Xiaojun Wan}.} \bibinfo{year}{2021}\natexlab{}.
\newblock \showarticletitle{Learning a Product Relevance Model from Click-Through Data in E-Commerce}. In \bibinfo{booktitle}{\emph{Proceedings of the Web Conference 2021}} \emph{(\bibinfo{series}{WWW ’21})}. \bibinfo{publisher}{ACM}, \bibinfo{pages}{2890–2899}.
\newblock
\href{https://doi.org/10.1145/3442381.3450129}{doi:\nolinkurl{10.1145/3442381.3450129}}


\bibitem[Ye et~al\mbox{.}(2023)]%
        {QUALITY}
\bibfield{author}{\bibinfo{person}{Chengcan Ye}, \bibinfo{person}{Ting Peng}, \bibinfo{person}{Tim Chang}, \bibinfo{person}{Zhiyi Zhou}, {and} \bibinfo{person}{Feng Wang}.} \bibinfo{year}{2023}\natexlab{}.
\newblock \showarticletitle{Query-aware Multi-modal based Ranking Relevance in Video Search}. In \bibinfo{booktitle}{\emph{Proceedings of the 2023 Conference on Empirical Methods in Natural Language Processing: Industry Track}}, \bibfield{editor}{\bibinfo{person}{Mingxuan Wang} {and} \bibinfo{person}{Imed Zitouni}} (Eds.). \bibinfo{publisher}{Association for Computational Linguistics}, \bibinfo{address}{Singapore}, \bibinfo{pages}{322--330}.
\newblock
\href{https://doi.org/10.18653/v1/2023.emnlp-industry.31}{doi:\nolinkurl{10.18653/v1/2023.emnlp-industry.31}}


\bibitem[Zhang et~al\mbox{.}(2025)]%
        {NoteLLM-2}
\bibfield{author}{\bibinfo{person}{Chao Zhang}, \bibinfo{person}{Haoxin Zhang}, \bibinfo{person}{Shiwei Wu}, \bibinfo{person}{Di Wu}, \bibinfo{person}{Tong Xu}, \bibinfo{person}{Xiangyu Zhao}, \bibinfo{person}{Yan Gao}, \bibinfo{person}{Yao Hu}, {and} \bibinfo{person}{Enhong Chen}.} \bibinfo{year}{2025}\natexlab{}.
\newblock \bibinfo{title}{NoteLLM-2: Multimodal Large Representation Models for Recommendation}.
\newblock
\showeprint[arxiv]{2405.16789}~[cs.IR]
\urldef\tempurl%
\url{https://arxiv.org/abs/2405.16789}
\showURL{%
\tempurl}


\bibitem[Zhang et~al\mbox{.}(2023)]%
        {Video-LLaMA}
\bibfield{author}{\bibinfo{person}{Hang Zhang}, \bibinfo{person}{Xin Li}, {and} \bibinfo{person}{Lidong Bing}.} \bibinfo{year}{2023}\natexlab{}.
\newblock \bibinfo{title}{Video-LLaMA: An Instruction-tuned Audio-Visual Language Model for Video Understanding}.
\newblock
\showeprint[arxiv]{2306.02858}~[cs.CL]
\urldef\tempurl%
\url{https://arxiv.org/abs/2306.02858}
\showURL{%
\tempurl}


\bibitem[Zhu et~al\mbox{.}(2023)]%
        {MiniGPT-4}
\bibfield{author}{\bibinfo{person}{Deyao Zhu}, \bibinfo{person}{Jun Chen}, \bibinfo{person}{Xiaoqian Shen}, \bibinfo{person}{Xiang Li}, {and} \bibinfo{person}{Mohamed Elhoseiny}.} \bibinfo{year}{2023}\natexlab{}.
\newblock \bibinfo{title}{MiniGPT-4: Enhancing Vision-Language Understanding with Advanced Large Language Models}.
\newblock
\showeprint[arxiv]{2304.10592}~[cs.CV]
\urldef\tempurl%
\url{https://arxiv.org/abs/2304.10592}
\showURL{%
\tempurl}


\bibitem[Zou et~al\mbox{.}(2021)]%
        {Pyramid-Ernie}
\bibfield{author}{\bibinfo{person}{Lixin Zou}, \bibinfo{person}{Shengqiang Zhang}, \bibinfo{person}{Hengyi Cai}, \bibinfo{person}{Dehong Ma}, \bibinfo{person}{Suqi Cheng}, \bibinfo{person}{Daiting Shi}, \bibinfo{person}{Zhifan Zhu}, \bibinfo{person}{Weiyue Su}, \bibinfo{person}{Shuaiqiang Wang}, \bibinfo{person}{Zhicong Cheng}, {and} \bibinfo{person}{Dawei Yin}.} \bibinfo{year}{2021}\natexlab{}.
\newblock \bibinfo{title}{Pre-trained Language Model based Ranking in Baidu Search}.
\newblock
\showeprint[arxiv]{2105.11108}~[cs.IR]
\urldef\tempurl%
\url{https://arxiv.org/abs/2105.11108}
\showURL{%
\tempurl}


\end{thebibliography}



\end{document}